\documentclass[3p,11pt,authoryear]{elsarticle}

\usepackage{amsfonts}
\usepackage{amsmath}
\usepackage{amssymb}
\usepackage{amsthm}
\usepackage{mathrsfs}
\usepackage{graphics}
\usepackage{lscape}
\usepackage{changebar}
\usepackage{graphicx}
\usepackage{epsfig}
\usepackage{bm}
\usepackage{color}
\newcommand{\tr}{\hbox{tr}}

\newcommand{\B}[1]{{\bm{#1}}}
\newcommand{\C}[1]{{\mathcal{#1}}}
\newcommand{\pa}{\partial}
\renewcommand{\vec}[1]{\mathbf{#1}}

\newcommand{\chib}{\mbox{\boldmath$\chi$}}

\newcommand{\deriv}[2]{\frac{\partial #1}{\partial #2}}

\journal{Journal of the Mechanics and Physics of Solids}

\begin{document}

\begin{frontmatter}

\title{Two-temperature continuum thermomechanics of deforming amorphous solids}

\author{Ken Kamrin$^1$ and Eran Bouchbinder$^2$}
\address{$^1$ Department of Mechanical Engineering, MIT, Cambridge, Massachusetts 01239, USA\\
$^2$ Chemical Physics Department, Weizmann Institute of Science, Rehovot 76100, Israel}
\date{\today}
\begin{abstract}
There is an ever-growing need for predictive models for the elasto-viscoplastic deformation of solids. Our goal in this paper is to incorporate recently developed out-of-equilibrium statistical concepts into a thermodynamically consistent, finite-deformation, continuum framework for deforming amorphous solids. The basic premise is that the configurational degrees of freedom of the material --- the part of the internal energy/entropy that corresponds to mechanically stable microscopic configurations --- are characterized by a configurational temperature that might differ from that of the vibrational degrees of freedom, which equilibrate rapidly with an external heat bath.
This results in an approximate internal energy decomposition into weakly interacting configurational and vibrational subsystems, which exchange energy following a Fourier-like law, leading to a thermomechanical framework permitting two well-defined temperatures. In this framework, internal variables that carry information about the state of the material equilibrate with the configurational subsystem, are explicitly associated with energy and entropy of their own, and couple to a viscoplastic flow rule. The coefficients that determine the rate of flow of entropy and heat between different internal systems are proposed to explicitly depend on the rate of irreversible deformation. As an application of this framework, we discuss two constitutive models for the response of glassy materials, a simple phenomenological model and a model related to the concept of Shear-Transformation-Zones as the basis for internal variables. The models account for several salient features of glassy deformation phenomenology. Directions for future investigation are briefly discussed.
\end{abstract}

\begin{keyword}

\end{keyword}

\end{frontmatter}

\section{Introduction: Background and Motivation}

The elasto-viscoplastic deformation of solids is a complex process that couples widely separate time and length scales. In polycrystalline solids, plastic deformation is mainly mediated by dislocations, which form complex and strongly interacting networks that give rise to a rich macroscopic phenomenology. In disordered solids (e.g. noncrystalline/amorphous solids below their glass temperature and granular materials), which lack long-range structural order, plastic deformation is mediated by localized shear-driven rearrangements involving a small number of basic elements (molecules/atoms/grains). These rearrangements produce inhomogeneous mesoscopic deformation fields that affect their surroundings. In both cases, the small scale dynamics are highly heterogenous, intermittent and seemingly non-deterministic, yet the emergent macroscopic phenomena are quite robust. Developing predictive continuum models that effectively bridge over the widely separate scales remains a major challenge of prime scientific and technological importance.

The most successful methodology for bridging over widely separated scales is equilibrium statistical thermodynamics. However, the viscoplastic deformation of solids is an intrinsically out-of-equilibrium phenomenon, hence standard equilibrium thermodynamics is not directly applicable in this case. Nevertheless, statistical concepts and the most general form of the laws of thermodynamics might still be applicable. Our goal in this paper is to incorporate recently developed out-of-equilibrium statistical concepts into a continuum thermomechanical framework for the large elasto-viscoplastic deformation of amorphous solids. Models for the elasto-viscoplastic deformation of amorphous solids have been quite extensively discussed in recent literature, see for example \citet{Sollich1997, Falk1998, Falk2011, AnandSu2005, Demetriou2006, Henann2008, Jiang2009, Homer2010, Abdeljawad2010, Li2013}.

The basic idea that has emerged in the statistical physics community in the last 15 years or so, is that the degrees of freedom of an amorphous/glassy material (i.e. the various contributions to the internal energy and entropy) can be approximately decomposed into two qualitatively different groups \citep{Cugliandolo1997, Berthier2000, Teff_review_Leuzzi, Teff_review_2011}. One group includes the vibrational motions of particles around a mechanically stable configuration (sometimes termed an inherent structure). These motions equilibrate with the thermal reservoir on a microscopic timescale; i.e. the temperature apparent in the vibrational fluctuations quickly evolves to match that of the reservoir. The other group includes the mechanically stable configurations themselves and the slow configurational rearrangements that allow transitions between them. These are mainly driven by external mechanical forces, possibly assisted by thermal fluctuations. It is the configurational degrees of freedom that fall out of internal equilibrium with the vibrational degrees when a glass is formed; and it is these degrees of freedom that undergo structural relaxation when a glass ages and evolve when a glass deforms plastically.

Numerous recent theoretical, numerical and experimental studies have provided quantitative evidence that the configurational degrees of freedom of either an aging or a deforming amorphous/glassy material attain a temperature of their own, which is markedly different from the temperature corresponding to the vibrational fluctuations of the microconstituents. The early works demonstrating the existence of a configurational temperature, usually termed an ``effective temperature'' in the statistical physics community, were rather exclusively based on deviations from the equilibrium fluctuation-dissipation relation (for a very recent and comprehensive review see \citet{Teff_review_2011}). That is, it was shown (mainly using simple models and molecular dynamics simulations) that the long time dynamics of an aging glassy material satisfy a generalized fluctuation-dissipation relation with a temperature that is different from the temperature associated with the vibrational motion of the microconstituents \citep{Cugliandolo1997, Teff_review_2011}. Of special importance for our present purposes are the colloidal glass experiments of \citet{Maggi2010}, which explicitly demonstrate that the effective temperature of a sample slowly evolves toward the thermal (i.e. vibrational) temperature over very long times, and eventually approaches it. This suggests that the two temperatures interact thermodynamically through an equilibration process akin to Fourier's relation to reach a common value. Generalized fluctuation-dissipation relations were also tested for steadily deforming simulated glasses, again demonstrating the existence of a well-defined effective temperature \citep{Berthier2002, Berthier2002a}.

Another line of investigation focused on ``athermal systems'', i.e. systems where no ordinary thermal (vibrational) fluctuations take place either due to coupling to a zero temperature bath (e.g. in computer simulations) or when the basic elements are too big to experience thermal vibrations. Realistic examples include foams (made of bubbles) and granular materials (made of grains) where the large size of the basic elements induces dissipation that rapidly quenches vibrations. In \citet{Ono2002}, Durian's foam model \citep{Durian1995} was studied under slow, steady shear deformation conditions. Five different definitions commonly used to define the ordinary thermal (vibrational) temperature --- pressure and energy fluctuation relations, the Green-Kubo relation for the shear viscosity, the Stokes-Einstein relation, and the thermodynamic definition from the rate of change of configurational energy with entropy --- were tested and it was discovered that all definitions yielded a similar effective temperature, though the system itself was strictly athermal. Similar ideas were developed and tested for granular materials \citep{Makse2002}. Both \citet{Makse2002} and \citet{Ono2002} showed by explicit calculation that the effective temperature is equivalently given by the derivative of the configurational energy with respect to the entropy of configurational states. This implies that the configurational subsystem behaves as a distinct thermodynamic entity, featuring a configurational temperature defined as the derivative of the configurational energy with respect to the configurational entropy, precisely as we formulate in detail below.

Experimental verification of the effective temperature concept is currently limited to systems in which the basic elements are sufficiently large to be directly tracked, such as grains in granular materials and colloidal particles in colloidal glasses \citep{GM2005, Colloids2006}. We finally note that the literature on the effective (configurational) temperature is immense and cannot be exhaustively reviewed here. The interested reader is referred to \citet{Teff_review_Leuzzi, Teff_review_2011} for additional information and relevant references.

Two-temperature thermodynamic frameworks for spatially homogeneous, non-plastically deforming, glassy materials have been investigated by several authors \citep{Nieuwenhuizen1998, Sciortino1999, Sciortino2000, Ottinger2006}, and extended models coupling the evolution of internal state variables to the new thermodynamic variables have begun to emerge. Applications to amorphous plasticity -- recent Shear-Transformation-Zone (STZ) models \citep{Langer2004, Falk_Teff_2007, BLI2009, BLII2009, BLIII2009, Falk2011, Gibou2012} --, to soft glassy rheology \citep{SollichCates2012}, to dislocation-mediated crystal plasticity \citep{Dislocations_Teff}, to glassy memory effects \citep{Kovacs_Teff_2010} and to hard-sphere materials \citep{Hard_spheres2012} have been proposed. Very recently, this framework has been used to explain a ductile-to-brittle transition in bulk metallic glasses \citep{Rycroft2012}.

With this background, we present the three major goals of this paper:

\begin{itemize}
\item To determine the general features of a rational thermomechanical framework which permits, at the outset, the existence of a second primitive temperature.  We view this as a question of basic, fundamental interest.
\\
\item To show that the essential two-temperature principles developed in the statistical physics community can be fit within a consistent thermomechanical framework.
\\
 \item To highlight certain non-trivial phenomena captured by the resulting constitutive framework, when choosing forms for various constitutive functions.
\end{itemize}

We proceed through several steps. 
In the first step, we describe the decomposition of the standard continuum laws of energy balance and entropy imbalance (cf. \cite{Truesdell_book}) into sub-balances for the two subsystems. At this point, we demonstrate basic thermomechanical consequences of these hypotheses (including items 1, 2 and 3 on the list below) by presenting a simple example of a two-temperature plasticity model. To develop a micro-statistically motivated model we next revert back to our basic hypotheses and, borrowing certain notions from \cite{BLII2009, BLIII2009, Bouchbinder2013}, introduce a minimal set of internal state variables, and permit the energy and entropy of the configuration to depend on them.  We assume defect-like functional forms for the entropy and energy of the internal variables, and assume constitutive dependences for the other variables consistent with the subsystems decomposition. Restrictions based on the second-law of thermodynamics then lead to a set of ``Fourier-type relations'' for heat transfer between the subsystems and through space, which give rise to a pair of heat equations, one for each temperature. Finally, we develop a specific plastic flow rule based on an STZ-micromechanism \citep{Falk1998}. The resulting model is then shown to capture a number of non-trivial phenomena observed in glassy deformation, including
\begin{enumerate}
\item Structural rejuvenation (i.e. increase in structural disorder) under shearing.
\item A finite configurational temperature in slow, steady shearing.
\item Slow aging of the structural degrees of freedom near the glass transition.
\item Newtonian flow near the glass transition.
\item Emergence of a flow stress for steady flows at low temperatures.
\end{enumerate}

\section{Basic hypotheses and the laws of thermodynamics}

Our goal in this section is to mathematically express the basic hypotheses discussed in the introduction and to systematically explore the implications of them for thermodynamics. This will set the stage for the thermomechanical analysis to be presented later.
\begin{figure}
\begin{center}
\epsfig{file=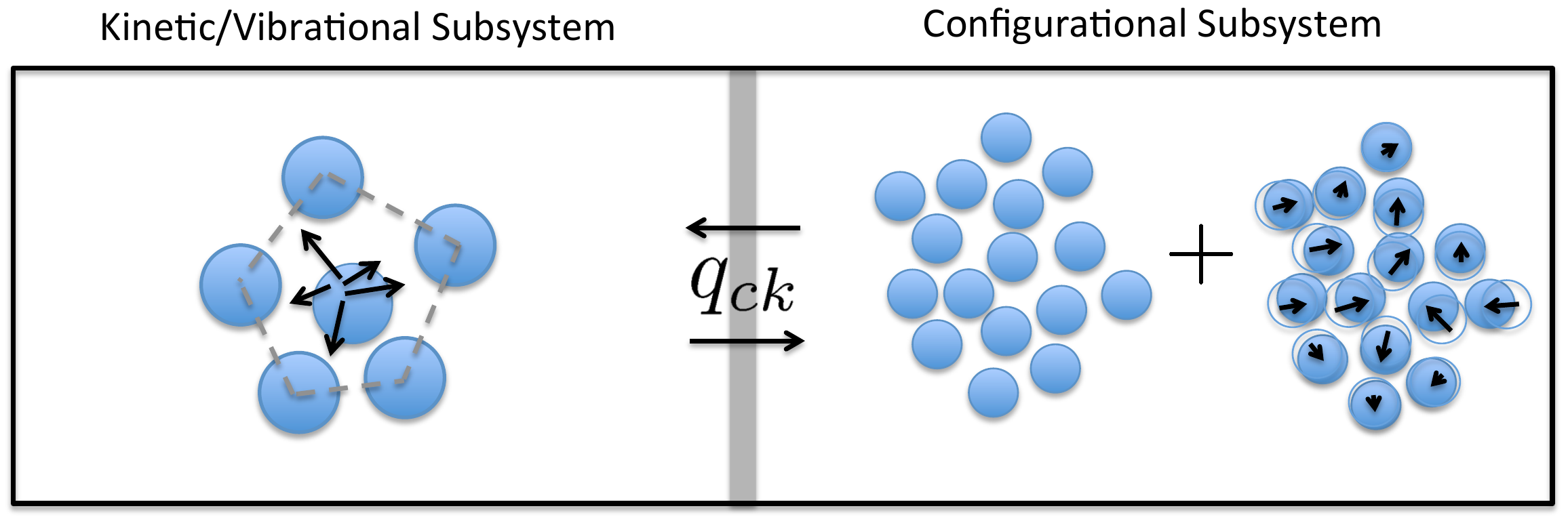, width=6in}
\caption{The ``two-block'' picture of the degrees of freedom in a general amorphous system: The kinetic-vibrational subsystem contains the energy contribution due to random thermal vibrations of particles in their cage of nearest neighbors (blown-up for clarity), and the configurational subsystem contains the bond energy of the mechanically stable configurations (inherent structures) and the energy of non-affine motion (with respect to the continuum-scale deformation) of the structure itself as it makes transitions from one stable configuration to another, averaged over the vibrational timescale. These motions correspond to local, infrequent, particles rearrangements accumulated over long timescales. A finite ``heat passage'' $q_{ck}$ can exist to transfer energy from whichever subsystem is ``hotter'' to the one that is ``colder''.} \label{Fig1}
\end{center}
\end{figure}

\subsection{Nonequilibrium subsystems of internal energy (and entropy): An analogy}

By ``nonequilibrium'' processes we refer to thermodynamic processes that do not immediately maximize the entropy. As a motivational example, consider a closed system containing two identical blocks of a highly heat-conductive material, which make contact on either side of a thermally conducting layer. The layer conducts heat between the blocks (labeled A and B) at a rate $q_{A B}\!=\!\kappa\ (\theta_A-\theta_B)$, where $\theta_A$ and $\theta_B$ are the blocks' temperatures. If $\kappa\to\infty$, the system equilibrates on a microscopic timescale, i.e. a temperature difference cannot exist for a finite time. If $\kappa\!=\!0$, we have a completely decoupled nonequilibrium system, for which the system is composed of two separate equilibrium subsystems that never approach a mutual equilibrium temperature (if $\theta_A\neq\theta_B$ initially). For finite $\kappa$, we obtain a non-trivial, nonequilibrium system: any temperature difference between the blocks gradually approaches an equilibrium temperature and the joint system gradually approaches a maximum entropy state.

As discussed above, this latter case is analogous to observations made in amorphous materials. Instead of two blocks separated in space, the material is viewed as having two weakly coupled subsystems appearing within the internal energy and entropy functions. Consider an idealized amorphous system containing randomly and densely packed interacting particles. Focus first on the \emph{kinetic/vibrational} (K/V) degrees of freedom, which describe random thermal vibrations around a mechanically stable microscopic configuration. In the amorphous setting, these degrees of freedom manifest as the part of the internal energy that arises due to the particles rattling about some mean positions within their cages of nearest neighbors. This subsystem rapidly equilibrates with an external bath. In the vicinity of the glass transition temperature and below it, the timescale of these vibrations is much faster than the slow timescale characterizing structural relaxation and typical rates of deformation. Indeed, the subsystems' separation is essentially a timescales separation. By averaging over the short vibrational timescale, the energy stored within the microscopic structure itself can be measured, which is classified as energy in the \emph{configurational} subsystem.  Moreover, as the structure is deformed into new configurations, we should also account in this subsystem for the energy due to non-affine\footnote{Throughout, we use the term ``non-affine'' to describe the part of the microscopic particle motion that differs from the continuum-level motion one obtains when averaging over representative volume elements of material.} motion of the particles (averaged over the vibrational timescale), resulting from the intermittent collective particle-level rearrangements. Under most circumstances, the kinetic energy of this movement comprises a very small contribution to the configurational energy. The \emph{internally nonequilibrium} nature of the material is characterized by the fact that for a given total internal energy, the distribution of energy into configurational and K/V subsystems does not generally maximize the total entropy. Rather, a finite energy exchange rate $q_{ck}$ can exist between the two subsystems, passing energy from the higher to the lower temperature subsystem, which plays the role of $q_{AB}$ in the previous illustrative example. Indeed, the internal heat flow $q_{ck}$ accounts for entropy production in the material, together with more conventional sources of entropy production, such as spatial heat fluxes.

The finiteness of $q_{ck}$ in the amorphous setting can be distinguished from the example cases of weakly sheared Newtonian fluids or elastically deformed crystals. In a weakly sheared fluid, thermal vibrations and stress-relaxing hopping motions occur on timescales that are not many orders of magnitude separated, and all of the degrees of freedom can be characterized by a single temperature. Hence, equipartition is satisfied, i.e. the total configurational energy remains in equilibrium with the kinetic degrees of freedom, analogous to $\kappa\!\to\!\infty$. In an elastically deformed crystal, both thermal vibrations and the external driving forces are, by the definition of the elastic regime, too weak to induce structural changes (e.g. dislocations motion), and hence the configurational degrees of freedom are essentially decoupled from the vibrational ones, analogous to $\kappa\!\to\!0$. Nevertheless, the microscopic kinetic and potential energies due to the crystal vibrations do maintain equilibrium following the equipartition theorem, which is completely analogous to our caricature of the K/V subsystem as single ``block'' in the thermodynamic analysis. In the case of shearing an amorphous solid, as was discussed in detail above (see, for example, \citet{Makse2002}), there exists ample evidence that the configurational degrees of freedom also form a single equilibrated ``block'', which is distinguished from the K/V ``block'', and hence two distinct temperatures may be sufficient to approximately describe the internal thermodynamics of the system. This physical picture is schematically sketched in Fig. \ref{Fig1}.

Let us denote by $\theta_c$  the temperature of the configurational subsystem and by $\theta_k$ the temperature of the kinetic/vibrational subsystem, which will both be formalized in detail in the following sections. The configurational temperature $\theta_c$ is a generalization of the relatively old notion of a fictive temperature \citep{Tool1946, Narayanaswamy1971}. It is typically {\em higher} than the thermal temperature $\theta_k$, because the structure typically retains memory of configurations characteristic of the higher temperature prior to a rapid quench and is also constantly being rejuvenated by the external driving forces. Following our two-block analogy, one would expect a Fourier-type relation of the form $q_{ck}=\kappa_{ck}\ (\theta_c-\theta_k)$. We note in passing that thermo-elastic effects (e.g.  thermal expansion, temperature-sensitive elastic moduli) provide another form of interaction between the two subsystems, which is not mediated by the heat flux $q_{ck}$. \emph{We neglect these phenomena in the present work} to focus more heavily on the deviatoric (non-volumetric) plastic response.

An important subtlety of the ensuing analysis pertains to the functional dependence of the conductivity $\kappa_{ck}$. In the absence of plastic deformation, only ordinary thermal fluctuations can pull the system out of a local configuration and induce flow from the configurational subsystem to the kinetic/vibrational one. As the glass transition is approached, thermal fluctuations become increasingly ineffective in spontaneously inducing structural relaxation, and consequently $\kappa_{ck}$ becomes extremely small in the absence of imposed shearing (it strictly vanishes if there exists an ideal glass transition). In the presence of plastic deformation, rearrangements are activated and energy can again flow between the two subsystems, i.e. $\kappa_{ck}\!>\!0$, at a rate that is affected by the rate of plastic deformation itself. Moreover, the rearrangements themselves can mediate diffusion of $\theta_c$, when the latter is spatially inhomogeneous. This occurs microscopically through the long range Eshelby-like fields that are generated by the localized shear rearrangements (that can be viewed as Eshelby-like inhomogeneities \citep{STZ_Eshelby_2004}). The direct dependence of the inter-subsystem conductivity and intra-configurational subsystem diffusivity on the rate of plastic deformation is a unique feature of these systems and hence of the corresponding material models. We now turn to formulate these ideas mathematically.

\subsection{Mathematical development}\label{math}

Consider a deformation process for which $\B {X}$ is the reference coordinate, $\B {x}$ is the deformed coordinate and $\B {x}\!=\!\B \chi(\B {X}, t)$, where $\B \chi$ is the motion function. The spatial velocity field is then defined by $\B {v}(\vec{x},t)=\dot{\B \chi}(\B \chi^{-1}(\B x,t),t)$, where $\dot{}$ denotes the Lagrangian time derivative. The local first law of thermodynamics reads
\begin{equation}
\label{first} \rho\,\dot{\epsilon}=\B T\!:\!\B L-\nabla\!\cdot\!\B q \ ,
\end{equation}
where $\rho$ is the mass density, $\epsilon$ is the internal energy per unit mass, $\B T$ is the Cauchy stress tensor, $\B L$ is the spatial velocity gradient (i.e $L_{ij}\!=\!\partial v_j/\partial x_i$), $\B q$ is the spatial heat flux and $\nabla\cdot$ is the spatial divergence operator. In the above we have assumed no internal heat sources or sinks, which would otherwise add a scalar heating term to the right hand side\footnote{This assumption is not crucial to the development; we could include a source/sink term, but its absence eases the derivation. Its inclusion would be needed if particle collisions are inelastic \citep{Campbell1990}, or in circumstances with true internal heat sources, e.g. radiation-rich environments.}. The second law of thermodynamics reads
\begin{equation}
\rho\,\dot{\eta}+\nabla\!\cdot\!\B j \ge 0 \label{second} \ ,
\end{equation}
where $\eta$ is the internal entropy per unit mass and $\B j$ the entropy flux.

In the spirit of the previous discussion, we now apply the assumption that the thermodynamical quantities can be approximately decomposed into their
kinetic/vibrational and configurational contributions, denoted hereafter by the subscripts $k$ and $c$ respectively.   First, we separate the internal energy and entropy into configurational and kinetic/vibrational contributions
\begin{eqnarray}
\epsilon \simeq \epsilon_c+\epsilon_k \quad\hbox{and}\quad\eta \simeq \eta_c+\eta_k  \ .
\end{eqnarray}
Next, the heat flux is assumed to approximately decompose as
\begin{equation}
\B q  \simeq \B q_c + \B q_k \ ,
\end{equation}
where $\B {q}_c$ and $\B {q}_k$ are the fluxes of configurational energy ($\epsilon_c$) and K/V energy ($\epsilon_k)$, respectively. To gain physical insight into this decomposition, note that the heat flux $\B q$ into (out of) a part of a body represents the flux of energy due to the flow of randomized movement of the microscopic constituents --- movements that collectively have zero mean velocity relative to the continuum-scale motion.  In our context, we can envision two different manifestations of the passage of randomized movement. The kinetic heat flux $\B q_k$ may be seen as the ``standard'' flux describing the passage of random vibrational motion into (out of)  a part. The configurational contribution $\B q_c$ can be understood as the energy flux due to rearrangements of the structure, which pass energy across space by disturbing the neighboring structural assembly in a random fashion, and which does not count any K/V vibrational mechanisms in so doing.

The entropy flux is similarly decomposed into two additive contributions
\begin{equation}
\B j \simeq \B j_c +\B j_k \ .
\end{equation}
In the classical Clausius-Duhem inequality, the temperature  $\theta$ is positted as a scalar, non-negative field, which relates the heat and entropy fluxes by $\B j \!=\! \B q/\theta$.  We introduce the notion of a second primitive temperature --- the most noteworthy departure from common thermomechanics in this paper --- by assuming analogously that each subsystem contribution to the entropy flux is proportional to its corresponding
heat flux via an as-yet-undefined temperature field, giving $\B j_c=\B q_c/\theta_c$ and $\B j_k=\B q_k/\theta_k$.  As a result
\begin{eqnarray}
\label{1st}
&&\rho\,\dot{\epsilon}=\B T\!:\!\B L-\nabla\!\cdot\!\B q_k-\nabla\!\cdot\!\B q_c \ ,\\
\label{2nd} &&\rho\,\dot{\eta}+\nabla\!\cdot\!(\B
q_k/\theta_k)+\nabla\!\cdot\!(\B q_c/\theta_c) \ge 0 \ .
\end{eqnarray}
The case of $\theta_k\ne\theta_c$  corresponds to internal nonequilibrium, whereas  $\theta_k\!=\!\theta_c\equiv\theta$ in internal equilibrium and we recover the
standard entropy flux $\B j \!=\! \B q/\theta$. We also assume each temperature is necessarily non-negative. We should also mention that our choice of $\B j_c=\B q_c/\theta_c$ and $\B j_k=\B q_k/\theta_k$ is somewhat analogous to multi-temperature thermomechanical models for immiscible mixtures \citep{Bowen1979, Dunwoody68}; instead of $c$ and $k$ subsystems, each phase of the mixture is permitted its own temperature and the relation between heat and entropy fluxes within each phase, and their additive appearance in the second law, appears in direct analogy to ours.

Adopting the conventional Kr\"{o}ner-Lee decomposition of the total
deformation gradient tensor $\B F$ \citep{kroner60},
\begin{equation}\label{KL}
\partial \B \chi/\partial\B {X}\equiv\B {F}={\B F}^e\B {F}^p,
\end{equation}
we can write the stress-power as $ \B T:\B L \!=\! J^{e-1}\B T^e:\dot{\B E}^e + J^{e-1}\B T^p:\B L^p$ with $J^e\!=\!\det \B F^e$, $\B {C}^e\!=\!\B {F}^{eT}\B {F}^e$,
$\B E^e\!=\!(\B {C}^e-\B {1})/2$,
$\B T^e\!=\!J^e\B {F}^{e-1}\B T\B {F}^{e-T}$,
$\B T^p\!=\!\B {C}^e\B T^e$ and
$\B L^p\!=\!\dot{\B {F}}^p\B {F}^{p-1}$ \citep{Gurtin10}. We note that $\B F^p$ maps the reference (undeformed) space into the so-called intermediate space and that $\B F^e$ maps the intermediate space into the deformed space. Also note that $J^p\!=\!\det\B F^p$ and hence $J\!=\!\det\B F\!=\!J^e J^p$.

In what follows we assume that plastic shear deformation involves no volumetric changes, i.e. we set $J^p\!=\!1$. In doing so we neither imply that there exists no coupling between plastic shear deformation and dilation/contraction nor that this coupling is of no importance. Such coupling is well documented for amorphous solids and might have consequences for their mechanics \citep{BMG_review_2007, Trexler2010}. We have developed a theory that includes pressure-sensitivity and shear dilation; however, the added complexities that arise suit reporting in a subsequent publication.

To proceed, we need separate energy balance equations for the two subsystems, which sum together to give the first law of thermodynamics in Eq. (\ref{1st}). We propose that these take the following form
\begin{align}
\label{firstk}
\rho\dot{\epsilon}_k&=\underbrace{J^{e-1}\B T^p_{dis}:\B L^p}_{\text{Plastic dissipation}}~+~\underbrace{q_{ck}}_{\text{Inter-subsystem heat flow}}~-~\underbrace{\nabla\cdot\B q_k}_{\text{Spatial kinetic heat flow}} \ ,\\
\label{firstc} \rho\dot{\epsilon}_c&=\underbrace{J^{e-1}\B T^e_{
}:\dot{\B E}^e}_{\text{Elastic power}}~+\underbrace{J^{e-1}\B T^p_{store}:\B L^p}_{\text{Plastic
storage power}}~-\underbrace{q_{ck}}_{\text{Inter-subsystem heat flow}}~-~\underbrace{\nabla\cdot\B q_c}_{\text{Spatial configurational heat flow}} \ ,
\end{align}

Let us discuss in detail the physical meaning of the different terms on the right-hand-sides of these equations. The spatial contributions to the heat flow {\it within} each subsystem are $\nabla\!\cdot\!\B q_k$ and $\nabla\cdot\B q_c$ for the kinetic and configurational subsystems respectively. $q_{ck}$ is the scalar internal heat flow that transfers energy {\it between} the configurational subsystem and the kinetic one. The rest of the terms represent the way the total stress-power $\B T\!:\!\B L$ is distributed among the subsystems. First, we note that elasticity resides in the configurational subsystem and hence reversible elastic energy appears in the $c$ subsystem energy balance; recall that we neglect thermo-elastic and volumetric effects, which could conceivably admit an elastic term in the $k$ subsystem. The remaining terms give the plastic stress-power $J^{e-1}\B T^p\!:\!\B L^p$. To remain general, we permit the plastic stress-power to affect both subsystems, by splitting $\B T^p$ into the sum of two contributions $\B T^p\!=\!\B T^p_{store}+\B T^p_{dis}$. The \emph{stored plastic power} describes the rate of stored energy of cold work, and thus contributes in the $c$ energy balance.  The \emph{plastic dissipation} is the rate of energy dissipated as ordinary heat, which is in the form of vibrations and thereby affects the $k$ subsystem balance.

Applying the chain rule to Eq. (\ref{2nd}) and substituting Eqs. (\ref{firstk})-(\ref{firstc}) into the result to eliminate
$\nabla\!\cdot\B q_k$ and $\nabla\!\cdot\B q_c$, we obtain
\begin{align}\label{int_space_1b}
\rho\dot{\eta}_c+\rho\dot{\eta}_k+&\frac{1}{\theta_c}\left(J^{e-1}(\B T^e:\dot{\B E}^e+\B
T^p_{store}:\B L^p) -q_{ck} -\rho\dot\epsilon_c\right)+\frac{1}{\theta_k}\left(J^{e-1}\B T^p_{dis}:\B L^p+q_{ck}-\rho\dot{\epsilon}_k\right)+\nonumber
\\
&-\frac{1}{\theta_k^2}\B q_k\cdot\nabla\theta_k
-\frac{1}{\theta_c^2}\B q_c\cdot\nabla\theta_c \ge 0 \ ,
\end{align}
where $\nabla$ is the spatial gradient operator. Letting  $\rho_I\!=\!J^e\rho$ be the mass density in the intermediate space, we redefine quantities relative to volumes, areas, and lengths in this space as
\begin{align}
\epsilon_c^I=\rho_I\epsilon_c \ \ , \ \ \epsilon_k^I=\rho_I\epsilon_k \ \ ,  \ \ \eta_c^I=\rho_I\eta_c \ \ , \ \ \eta_k^I=\rho_I\eta_k \ \ , \ \ q_{ck}^I=J^e q_{ck} \ \ ,
\nonumber\\
\B q_c^I=J^e\B F^{e-1}\B q_c \ \ ,   \ \  \B q_k^I=J^e\B F^{e-1}\B q_k \ \ ,  \ \ \B g_c^I=\B F^{eT}\nabla \theta_c \ \ ,  \ \ \B g_k^I=\B F^{eT}\nabla \theta_k \ .
\end{align}
Multiplying Eq. (\ref{int_space_1b}) by $J^e$ and substituting the above quantities, we obtain an intermediate frame entropy imbalance (for ease of notation, the $I's$ are omitted hereafter)
\begin{align}\label{int_space_2}
\dot{\eta}_c+\dot{\eta}_k+&\frac{1}{\theta_c}(\B T^e\!:\!\dot{\B E}^e+\B
T^p_{store}\!:\!\B L^p -q_{ck} -
\dot\epsilon_c)+\frac{1}{\theta_k}(\B T^p_{dis}\!:\!\B L^p+q_{ck}-\dot{\epsilon}_k)
-\frac{1}{\theta_k^2}\B q_k\cdot\B {g}_k
-\frac{1}{\theta_c^2}\B q_c\cdot\B {g}_c \ge 0  \ .
\end{align}
Similarly, the energy balances become
\begin{align}
\label{firstkI}
\dot{\epsilon}_k
&={\B T^p_{dis}:\B L^p}+q_{ck}-\text{Div}_I\B q_k \ ,\\
\label{firstcI} \dot{\epsilon}_c
&={\B T^e_{
}:\dot{\B E}^e}+{\B T^p_{store}:\B L^p}-q_{ck}-\text{Div}_I\B q_c \ ,
\end{align}
where $\text{Div}_I(\B u)\!\equiv\!J^e\nabla\cdot(J^{e-1}\B F^e \B u)$ for any vector field $\B u$.

\section{An example: A simple two-temperature plasticity model}
\label{sec:conventional}

As was stated in the abstract, internal variables that carry information about the structural state of the deforming solid, and determine its subsequent response, play a central role in a theory of elasto-plasticity. Yet, we present here first a simple two-temperature model for rate-dependent elasto-plasticity, which does not include internal variables. Our goal is to illustrate general features of the two-temperature thermomechanical framework.

We start by defining the configurational and kinetic Helmholtz free-energy per unit intermediate volume
\begin{align}
\psi_c&=\epsilon_c-\theta_c\eta_c, \ \ \  \psi_k=\epsilon_k-\theta_k\eta_k \ .
\end{align}
Let $\vec{D}^p\!=\!\text{sym}\ \vec{L}^p$ be the plastic rate-of-deformation tensor, $d^p\!=\!|\vec{D}^p|\!=\!\sqrt{\vec{D}^p:\vec{D}^p}$ the scalar plastic flow-rate, and $\vec{N}^p=\vec{D}^p/|\vec{D}^p|$ the direction of plastic flow.  Let us assume plastic incompressibility, $\tr \vec{L}^p=0$. We write the following functional dependences (note that here, and throughout the paper, we use the \ $\hat{}$  \ symbol to denote constitutive functions of designated independent variables)
\begin{align}
& \psi_c=\hat{\psi}_c(\vec{E}^e,\theta_c), \ \ \epsilon_c=\hat{\epsilon}_c(\vec{E}^e,\theta_c), \ \ \eta_c=\hat{\eta}_c(\theta_c), \ \ \psi_k=\hat{\psi}_k(\theta_k), \ \ \epsilon_k=\hat{\epsilon}_k(\theta_k), \ \ \eta_k=\hat{\eta}_k(\theta_k) \ ,
\label{conventional_cons1}\\
&\vec{T}^{p\prime}_{dis}=\hat{\vec{S}}^p(d^p,\vec{N}^p,\theta_c,\theta_k), \  \ \ \vec{T}^p_{store}=\hat{\vec{T}}^p_{store}(\vec{N}^p,\theta_k, \theta_c),  \  \ \ \vec{T}^e=\hat{\vec{T}}^e(\vec{E}^e) \ ,\label{conventional_cons2}\\
& q_{ck}=\hat{q}_{ck}(\theta_k,\theta_c,d^p) , \ \ \ \B q_c=\hat{\B q}_c(\theta_c,\B g_c , d^p), \ \ \ \B q_k=\hat{\B q}_k(\theta_k,\B g_k) \ .
\label{conventional_cons3}
\end{align}
We use $'$ to denote the deviator, and note that the forms for $\vec{T}^{p\prime}_{dis}$ and $\vec{T}^p_{store}$ assume $d^p\!>\!0$. These constitutive dependences are not unique. Rather, they are informed by past experience with constitutive theories of thermal plasticity, and are properly extended for consistency with the two-subsystem decomposition. In addition, the $\vec{T}_{dis}^p$ and $\vec{T}^{p\prime}_{store}$ dependences are also motivated by the statistical-thermodynamics-based model to be described in the upcoming sections. The above dependences are frame-indifferent (see Appendix \ref{frame_indifference}), and unlike standard heat transfer theories, we allow all heat flow variables associated with the configurational subsystem to explicitly depend on
the plastic rate of deformation. The latter is a key point, as plastic deformation directly influences the time-scale for configurational energy transfer. Finally, in the spirit of the subsystems separation, the above constitutive dependences aim at coupling configurational/kinetic dependent variables with configurational/kinetic independent variables.

Substituting the free-energies into the inequality in (\ref{int_space_2}) and the expanding time-derivatives, we obtain
\begin{align}
& -\frac{1}{\theta_c}\left(\deriv{\hat{\psi}_c}{\theta_c}+\eta_k\right)\dot{\theta}_c-\frac{1}{\theta_k}\left(\deriv{\hat{\psi}_k}{\theta_k}+\eta_c\right)\dot{\theta}_k+\frac{1}{\theta_c}\left(-\deriv{\hat{\psi}_c}{\vec{E}^e}+\vec{T}^e\right):\dot{\vec{E}^e} \nonumber
\\
&+\left(\frac{1}{\theta_c}\vec{T}^p_{store} +\frac{1}{\theta_k}\vec{T}^p_{dis}\right):\vec{D}^p-\frac{1}{\theta_k^2}\B q_k\cdot\B {g}_k
-\frac{1}{\theta_c^2}\B q_c\cdot\B {g}_c+\left(\frac{1}{\theta_k}-\frac{1}{\theta_c}\right)q_{ck}\ge 0 \ .
\label{simple_ineq}
\end{align}
As the first three terms are independent of $\dot{\vec{E}}^e$, $\dot{\theta}_k$ and $\dot{\theta}_c$, they must vanish in order to ensure non-violation of the inequality \citep{ColemanNoll1963}. Hence,
\begin{equation}
\vec{T}^e=\deriv{\hat{\psi}_c}{\vec{E}^e}, \ \ \eta_c=-\deriv{\hat{\psi}_c}{\theta_c}, \ \  \eta_k=-\deriv{\hat{\psi}_k}{\theta_k} \ ,
\label{conventional_Maxwell}
\end{equation}
which represent an elasticity relation and Helmholtz relations for each subsystem, respectively. Since $\eta_c$ is assumed independent of $\vec{E}^e$, the configurational Helmholtz relation implies that $\psi_c\!=\!\hat{\psi}_{c1}(\vec{E}^e)\!+\!\hat{\psi}_{c2}(\theta_c)$.

To ensure that the four last terms in (\ref{simple_ineq}) do not violate the inequality, we assume -- as a sufficient, but not necessary condition -- that each is non-negative separately. For the last three terms, non-negativity is assured by the following set of Fourier-type relations
\begin{align}
{q}_{ck}=\kappa_{ck}\ (\theta_c-\theta_k) \ ,& \ \ \ \ \kappa_{ck}=\hat\kappa_{ck}(\theta_c, \theta_k, d^p)\ge 0,
\nonumber\\
\B q_c=-\kappa_c\ \B g_c\ ,& \ \ \ \ \kappa_c=\hat{\kappa}_c(\theta_c,d^p)\ge0,
\nonumber\\
\B q_k=-\kappa_k\ \B g_k\ ,& \ \ \ \ \kappa_k=\hat{\kappa}_k(\theta_k)\ge0.
\label{Fourier_conventional}
\end{align}
The term proportional to $\vec{D}^p$ can be made non-negative under various selections of the functions for $\vec{T}^{p\prime}_{dis}$ and $\vec{T}^p_{store}$. To obtain a simple, isotropic, incompressible Bingham-like rheology, we choose
\begin{align}
&\vec{T}^{p}_{store}=Y^p_{store}\vec{N}^p \ , \ \ \ Y^p_{store}=\hat{Y}^p_{store}(\theta_k, \theta_c)\ge 0,
\label{conventional_Tp}
\\
&\vec{T}^{p\prime}_{dis}={\mu^p} d^p\vec{N}^p \ ,  \ \ \ \ \  \mu^p=\hat{\mu}^p(\theta_c,\theta_k)\ge 0 \ ,
\label{conventional_flow}
\end{align}
which is valid only for $d^p\!>\!0$. Together, the plastic stress obeys $\vec{T}^{p \prime}=\hat{Y}^p_{store}(\theta_k, \theta_c)\vec{N}^p + \hat{\mu}^p(\theta_c,\theta_k) d^p\vec{N}^p$, which makes it evident that $\mu^p$ is an effective plastic flow viscosity and $Y_{store}^p$ acts as a yield strength.

Finally, we insert these results into Eqs. (\ref{firstkI}) and (\ref{firstcI}) to obtain heat equations for the two subsystems. Applying the chain rule and the definition of the free-energies, we can write $\dot{\epsilon}_k\!=\!\deriv{\hat{\epsilon}_k}{\theta_k}\dot{\theta}_k\!\equiv\!\hat{c}_k(\theta_k)\dot{\theta}_k$ and $\dot{\epsilon}_c\!=\!\deriv{\hat{\epsilon}_c}{\theta_c}\dot{\theta}_c\!+\!\deriv{\hat{\psi}_c}{\vec{E}^e}\dot{\vec{E}}^e\!\equiv\!\hat{c}_c(\theta_c)\dot{\theta}_c\!+\!\deriv{\hat{\psi}_c}{\vec{E}^e}\dot{\vec{E}}^e$, to obtain
\begin{align}
c_k \dot\theta_k =& \
 \mu^p d^{p2}
+{\kappa}_{ck}\ (\theta_c-\theta_k)+ \text{Div}_I\left[\kappa_k \,\B g_k\right] \ ,
\label{conventional_heat_k}
\\
c_c \dot\theta_c =& \ Y^p_{store} d^p
-{\kappa}_{ck}\ (\theta_c-\theta_k) +
\text{Div}_I \left[{\kappa}_c\,  \B g_c
\right] \ .
\label{conventional_heat_c}
\end{align}

To appreciate the implications of the two-temperature framework in general, and of the configurational heat equation in particular, we need to be more specific about how $\theta_c$ affects the flow rule. That is, we need to specify the $\theta_c$-dependence of $\hat{\mu}^p(\theta_c,\theta_k)$. In the subsequent sections, where we will discuss an internal variables model within a statistical thermodynamic framework, we will show that $\vec{D}^p$ is proportional to a Boltzmann-like factor with $\theta_c$ appearing as the relevant temperature. For the present discussion this implies that
\begin{equation}
\hat{\mu}^p(\theta_c,\theta_k) \propto e^{\frac{e_z}{k_B \theta_c}} \ ,
\label{mu_p}
\end{equation}
where $k_B$ is Boltzmann's constant and $e_z$ is a constant whose physical meaning will be explained later on. The most important thing for our purposes here is that Eq. (\ref{mu_p}) implies that the plastic rate of deformation $\vec{D}^p$ in Eq. (\ref{conventional_flow}) is very (i.e. exponentially) sensitive to variations in $\theta_c$. We therefore focus here on the behavior of $\theta_c$, according to Eq. (\ref{conventional_heat_c}), under various thermomechanical conditions:
\begin{itemize}
\item {\em Thermodynamic aging} -- Let us consider a situation in which a homogeneous sample is held at a temperature $\theta_k$ near its glass temperature with no imposed deformation, $d^p\!=\!0$. The evolution of the system under these conditions corresponds to ``structural relaxation'' or ``aging''.  Structural relaxation is a central glassy phenomenon in which physical properties (e.g. volume, enthalpy, elastic moduli etc.) vary spontaneously (i.e. in the absence of external driving forces) and slowly with no characteristic timescale \citep{Angell1988, Gotze1992, Angell2000, Cavagna2009}. In many cases, the time evolution is logarithmic. Aging is one of the most basic manifestations of the nonequilibrium nature of the glassy state. Under the stated conditions Eq. (\ref{conventional_heat_c}) predicts that
    \begin{align}
    \hat{c}_c \dot\theta_c = -\hat{\kappa}_{ck}(\theta_k,\theta_c)\ (\theta_c-\theta_k) \ .\label{conventional_aging}
    \end{align}
    Hence, our model predicts that $\theta_c$ evolves from its initial value to $\theta_k$; the two subsystems gradually equilibrate, and the $\theta_c$ relaxation dynamics correspond to structural aging in our nonequilibrium thermodynamic framework  (typically the initial condition satisfies $\theta_c\!>\!\theta_k$ and consequently $\theta_c$ decreases during the aging process). Since $\kappa_{ck}/c_c$ typically varies slowly and nonlinearly with $\theta_c$, the relaxation of $\theta_c$ to equilibrium at $\theta_k$ will be slow and strongly non-exponential.

    The thermodynamic picture of aging is partially supported by various generalized fluctuation-dissipation analyses of aging systems \citep{Teff_review_2011}. As was mentioned in the introduction, the recent colloidal glass experiments of \citet{Maggi2010} explicitly demonstrated that $\theta_c$ -- estimated through a generalized fluctuation-dissipation relation -- equilibrates with $\theta_k$ for sufficiently long aging times (cf. Fig. 3b there). While these observations provide strong support for the existence and importance of $\theta_c$ in aging dynamics, it is important to note, however, that even if the function representing $\kappa_{ck}/{c}_c$ is chosen so as to quantitatively match experimental aging data, this single-variable description of aging remains incomplete, cf. \citet{Kolvin2012}.

\item {\em Low-temperature homogeneous steady state deformation} -- Consider then homogeneous steady state deformation at a low temperature $\theta_k$. Under these conditions we can set $\dot\theta_c\!=\!0$ in Eq. (\ref{conventional_heat_c}) (due to the steady state conditions), neglect $\theta_k$ as compared to $\theta_c$ (low $\theta_k$) and omit the spatial term (homogeneous deformation). With these we can isolate $\theta_c$ in Eq. (\ref{conventional_heat_c}), obtaining
    \begin{equation}
    \theta_c = \frac{Y^p_{store} d^p}{\kappa_{ck}} > 0 \ .
    \label{conventional_thetac}
    \end{equation}
    This result shows that the configurational temperature $\theta_c$ attains a finite value (obtained by solving the above equation if $Y^p_{store}$ and $\kappa_{ck}$ depend on $\theta_c$) due to a competition between a ``rejuvenation'' term $Y^p_{store} d^p$ (acting to increase $\theta_c$) and a ``relaxation'' term $\kappa_{ck} \theta_c$ (acting to decrease $\theta_c$).
    As discussed above, at the low temperatures $\theta_k$ of interest here, it is sensible to assume that $\kappa_{ck}\!\propto\!d^p$. This implies that the steady state value of $\theta_c$ in Eq. (\ref{conventional_thetac}) is {\em independent} of the deformation rate (if $Y^p_{store}$ is independent of $d^p$), as is indeed widely observed (see, for example, \cite{Ono2002}). 
    
\item {\em Configurational temperature diffusion} -- The configurational heat diffusion term $\text{Div}_I \left[\kappa_c\ \B g_c \right]$ in Eq. (\ref{conventional_heat_c}) is expected to play a role in the presence of inhomogeneous flows and localization. In particular, $\kappa_c$ is inevitably related to a {\it length-scale} and is expected to give rise to, and potentially explain, shear-band diffusive widening. In this regard, it is possible that $\theta_c$ is related to the so-called ``fluidity'' state variable recently invoked in nonlocal flow theories \citep{Bocquet2009, Kamrin2012, Henann2013}. Like $\theta_c$, the fluidity is hypothesized to obey diffusional dynamics, which enable it to describe shear band widths in various amorphous media.

    In more general terms, as the configurational heat conductivity $\kappa_c$ introduces a length-scale, the two-temperature framework might provide a new mechanism for size effects in constitutive models. Unlike other size-dependent plasticity models where scale-sensitive plastic flow properties are obtained by introducing gradient-based pairings in the internal/external power statements \citep{fleck2001,gurtin2005}, the nonlocal behavior here emerges from the splitting of the usual first law into two pieces.  The diffusion of the new temperature $\theta_c$ then affects the flow through $\hat{\mu}^p(\theta_c,\theta_k)$. This coupling gives rise to a physically nonlocal constitutive mechanism, on time/space scales distinct from the usual (i.e. K/V) heat diffusion. This nonlocality bears a qualitative similarity to the implicit gradient model of \cite{anand2011}, where an additional diffusive variable (the ``fictitious plastic strain'') is introduced, and nonlocal plastic flow behaviors are induced by dependence of the flow rule on this variable.
\end{itemize}

\section{Internal variables and constitutive dependences}
\label{constit_sys}

It is our goal now to go beyond the illustrative simple model discussed in the previous Section and to derive a new constitutive model that integrates coarse-grained internal variables within the proposed framework. In so doing we hope to develop a more basic statistical thermodynamic approach, leading to models with improved predictive powers. Internal variables implicitly carry, through their evolution, information about the history of deformation, and, as descriptors of the material state, directly influence the deformation behavior \citep{Coleman1967, Rice1971, Maugin1994}.
The evolution of these variables, we posit, is driven by the exchange of energy and entropy with other parts of the deforming system and hence the internal variables -- which are naturally associated with the configurational subsystem -- must be explicitly associated with energy and entropy of their own \citep{BLII2009, Berdichevsky2008, Berdichevsky2012}. This is a rather significant deviation from conventional approaches, which traditionally do not associate entropy with internal variables \citep{Coleman1967, Rice1971, Maugin1994}.

In an ideal scenario, one should be able to define macroscopic internal variables by systematically coarse-graining over microscopic dynamics. In practice, this is a daunting task and one usually resorts to various phenomenological descriptions, which are sometimes rationalized based on physical intuition, experimental observations, symmetries, computer simulations, etc. An alternative approach would be to first identify mesoscopic objects that are abstracted from the microscopic physics and only then to coarse-grain to obtain the internal variables and their evolution, as in Shear-Transformation-Zone (STZ)-based theories \citep{Argon1979, Falk1998,Falk2011} to be discussed later. Before we specify such a microscopic mechanism (see Section \ref{STZ_sec}), our goal in this Section is to develop a broad theoretical framework that permits a minimal set of internal variables, to which a more specific micro-mechanism may later be attributed.

In amorphous solids, different initial states of disorder would result from different cooling rates through which the glass transition was approached. The different initial structural states may determine whether the stress-strain curve exhibits a stress peak or not, and whether the material shear localizes or not (see the review papers of \citet{BMG_review_2007, Trexler2010}). In addition, we know that solids (both crystalline and amorphous) carry orientational memory of previous plastic deformation, which affects subsequent deformation. This is demonstrated, for example, in the classical Bauschinger effect in metals and glassy polymers, where specimens undergo a decrease in flow strength in compression after experiencing a certain amount of plastic extension. To address these observations in a minimal sense, in what follows we consider one scalar and one (symmetric, deviatoric) tensorial internal variable, denoted respectively by $\Lambda$ and $\B M$. These variable types are common to isotropic and kinematic hardening models \citep{frederick2007}. In our setting, we assume that the internal variables $\Lambda$ and $\B M$ arise from dilute, weakly interacting flow defects (whose precise nature is not specified yet) that mediate plastic deformation. The internal variable $\Lambda$ stands for their density (number per unit intermediate volume) and the deviatoric tensor $\B M$ describes their (dimensionless) average orientation in the intermediate space.  Defining  $m\equiv|\B M|\!=\!\sqrt{\B M: \B M}$, we require $0\le m\le1$ where $m\!=\!0$ represents an isotropic orientational state and $m\!=\!1$ a completely anisotropic state where all defects align in the same direction.

We use the set $\{\epsilon_{k},\epsilon_c, \B E^e\}$, its time derivatives, and the set $\{ \Lambda, \B M, \B g_{k}, \B g_{c}, \B L^p \}$ as the independent variables, and assume the following constitutive dependences
\begin{eqnarray}\label{ini_const_deps}
\eta_k\!=\!\hat{\eta}_k(\epsilon_k), \quad \eta_c \!=\!\hat{\eta}_c(\epsilon_c,\B E^e, \Lambda, \B M), \quad \B T^e\!=\!\hat{\B T}^e(\B E^e)\ .
\end{eqnarray}
We stress that we use entropies and internal energies, rather than free-energies, as the basic quantities because we insist on directly associating entropy and energy with the internal variables. In our view, this constitutes an important deviation from conventional modelling. The above constitutive choices reflect the separation between subsystems by associating each entropy only to the variables naturally arising in that subsystem, and are also frame-indifferent (see Appendix \ref{frame_indifference}). We now follow, nearly verbatim, the steps described in Section \ref{sec:conventional}. Expanding time-derivatives in Eq. (\ref{int_space_2}), we obtain
\begin{align}
\label{second2}
&\left(\deriv{\hat{\eta}_k}{\epsilon_k}-\frac{1}{\theta_k}\right)\dot{\epsilon}_k+\left(\deriv{\hat{\eta}_c}{\epsilon_c}-\frac{1}{\theta_c}\right)\dot{\epsilon}_c+
\left(\frac{1}{\theta_c}{\B T}^e+\deriv{\hat{\eta}_c}{\B E^e}\right):\dot{\B E}^e\nonumber
\\
&
+\deriv{\hat{\eta}_c}{\B M}:\dot{\B M}+\left(
\frac{1}{\theta_c}\B T^p_{store}+\frac{1}{\theta_k}\B T^p_{dis}\right):\B L^p\nonumber
\\
&
-\frac{1}{\theta_k^2}\B q_k\cdot\B g_k
-\frac{1}{\theta_c^2}\B q_c\cdot\B g_c +\left(\frac{1}{\theta_k}-\frac{1}{\theta_c}\right)q_{ck}+\deriv{\hat{\eta}_c}{\Lambda}\dot{\Lambda}\ge 0 \ .
\end{align}

In spite of the fact that we do not yet specify explicit constitutive dependences for the other dependent variables in (\ref{second2}), let us assume they are all independent of $\dot{\epsilon}_c, \dot{\epsilon}_k,  \dot{\B E}^e$. Then, since the three first terms in the above inequality appear linear in the independent variables $\dot\epsilon_c$, $\dot\epsilon_k$, and $\dot{\B E}^e$ respectively, the corresponding coefficients must vanish to ensure non-violation of the inequality \citep{ColemanNoll1963}. Consequently,
\begin{equation}
\label{good_CN}
\hat{\B T}^e=-\theta_c\deriv{\hat{\eta}_c}{\B E^e},\quad
\frac{1}{\theta}_c=\deriv{\hat{\eta}_c}{\epsilon_c},\quad
\frac{1}{\theta_k}=\deriv{\hat{\eta}_k}{\epsilon_k} \ ,
\end{equation}
which are the counterparts of Eq. (\ref{conventional_Maxwell}), just expressed in terms of entropies and internal energies, rather than free-energies. Now we go beyond the simple model of Section \ref{sec:conventional}, and similar to \cite{BLII2009, BLIII2009} we express the entropy as
\begin{align}\label{ini_eta_split}
\hat{\eta}_c(\epsilon_c,\B E^e, \Lambda, \B M)=&
 \hat{\eta}_{d}(\Lambda,\B M) + \hat{\eta}_{sur}(\underbrace{\epsilon_c-\hat\epsilon_d(\Lambda,\B M)-\hat{\epsilon}_{e}(\B E^e)}_{\equiv\epsilon_{sur}}) \ .
\end{align}
Here, $\epsilon_d$ represents the part of the configurational energy corresponding to the energies of formation of the flow defects and $\epsilon_e$ is the energy from bulk elasticity.  Hence, $\epsilon_{sur}$ and $\eta_{sur}$ are the energy and entropy of the large background structure \emph{surrounding} the defects, excluding the elastic energy.
The $sur$ degrees of freedom satisfy $\partial\hat\eta_{sur}/\partial\epsilon_{sur}\!=\!\partial\hat\eta_c/\partial\epsilon_c\!=\!1/\theta_c$, and thus equivalently behave as the ``bath'' for $d$.  Also, under persistent driving, $d$ is expected to equilibrate with $sur$ relatively quickly compared to the (extremely) slow  equilibration between the $k$ and $c$ subsystems.

We choose $\hat{\epsilon}_e(\B E^e)\!=\!G |\B E^{e\prime}|^2\!+\!(\kappa/2)(\text{tr} \B E^e)^2$ such that Eqs. (\ref{good_CN})-(\ref{ini_eta_split}) give rise to the following elasticity relation
\begin{equation}\label{lin_elast}
\B T^e = -\theta_c\deriv{\hat{\eta}_c}{\B E^e}=\deriv{\hat{\epsilon}_e}{\B E^e}=2 G \B E^{e \prime}+\kappa (\text{tr}\B E^e)\mathbf{1}.
\end{equation}
Consequently, $\B T^p \!=\! \B C^e \B T^e$ is \emph{symmetric}, and $\B T^p_{store}$ and $\B T^p_{dis}$ are assumed symmetric as well.

We propose that the energy and entropy of the flow defects take the following forms
\begin{equation}
\epsilon_d = \hat\epsilon_d(\Lambda, \B M) = e_z \Lambda  \ , \quad\quad \eta_d =\hat\eta_d(\Lambda, \B M) = k_B\left[\Lambda - \Lambda\log\left(v_0 \Lambda\right) + \Lambda \hat{s}(m)\right] \ ,
\end{equation}
where $e_z$ is the formation energy of an isolated defect and $v_0$ is the characteristic volume of a defect. The entropy density of defects, $\hat{\eta}_d$, can be obtained in the dilute limit by counting the number of ways defects can be placed on available sites and then given an orientation. The dimensionless entropic factor $\hat{s}(m)$ can be modeled as follows.  Clearly $\hat{s}(1)\!=\!0$ as there is only one way to align defects to obtain $m\!=\!1$, modulo rigid rotation.  It should also approach a maximum value at $m\!=\!0$, the isotropic state, with $d\hat{s}/dm\!=\!0$ at $m\!=\!0$ due to symmetry. We therefore assume
\begin{equation}
\label{m_entropy}
\hat{s}(m)=(1-m^2)s_0 \ ,
\end{equation}
where $k_Bs_0$ is the orientational entropy per flow defect when $m\!=\!0$.
We can now write
\begin{equation}\label{etacderivs}
\frac{\pa \hat\eta_c}{\pa \Lambda} =
k_B[(1-m^2)s_0-\log\left(v_0\Lambda\right)]-\frac{e_z}{\theta_c}\ , \quad\quad \frac{\pa
\hat\eta_c}{\pa \B M} = -2 k_B
\Lambda s_0 {\B M}\ .
\end{equation}
Using these with Eqs. (\ref{good_CN}) in the second law (\ref{second2}) yields the following reduced second-law inequality
\begin{eqnarray}
\label{second3}
&&\left(-2 k_B s_0
\Lambda {\B M}\right):{\dot{\B M}}+\left(
\frac{1}{\theta_c}\B T^p_{store}+\frac{1}{\theta_k}\B T^p_{dis}\right):\B L^p
\\
&&-\frac{1}{\theta_k^2}\B q_k\cdot\B g_k
-\frac{1}{\theta_c^2}\B q_c\cdot\B g_c +\left(\frac{1}{\theta_k}-\frac{1}{\theta_c}\right)q_{ck}+\left(-\frac{e_z}{\theta_c}+k_B\left[(1-m^2)s_0-\log\left(v_0\Lambda\right)\right]\right){\dot{\Lambda}}\ge 0 \ .\nonumber
\end{eqnarray}

Let us now switch to using $\theta_{c}$ and $\theta_k$, instead of $\epsilon_{c}$ and $\epsilon_{k}$, as
independent variables for the remaining quantities and assume the following constitutive dependences
\begin{eqnarray}
&&q_{ck}=\hat{q}_{ck}(\theta_k,\theta_c,d^p) , \quad \B q_c=\hat{\B q}_c(\theta_c,\B g_c , d^p), \quad \B q_k=\hat{\B q}_k(\theta_k,\B g_k) \ ,\label{heat_fluxes}
\\
&&\B T^p_{dis}=\hat{\B T}^p_{dis}(\B D^p, \theta_c,\theta_k,\Lambda,\B M), \quad \B T^p_{store}=\hat{\B T}^p_{store}(\theta_c,\theta_k,\Lambda,\B M)
\\
&&\dot \Lambda=\hat{\dot{{\Lambda}}}(\B M, \Lambda,\theta_c,\B D^p), \quad
\dot{\B M}=\hat{\dot{{\B M}}}( \B M, \Lambda,\theta_c,\B D^p) \ ,\label{internal_constit}
\end{eqnarray}
which should be compared to Eqs. (\ref{conventional_cons2})-(\ref{conventional_cons3}).

For consistency with the new independent variables, we invert Eqs. (\ref{ini_const_deps}) and (\ref{ini_eta_split}) to obtain
\begin{equation}
\epsilon_k\!=\!\hat\epsilon_k(\theta_k), \quad \epsilon_c\!=\!\hat{\epsilon}_c(\theta_c,\B E^e,\Lambda,\B M)\!=\!\hat{\epsilon}_{sur}(\theta_c)+\hat{\epsilon}_d(\Lambda,\B M)+\hat{\epsilon}_{e}(\B E^e).
\end{equation}
The above gives important perspective on how the different variables influence different contributions to the total internal energy. The configurational energy decomposes into the energy of the structure outside the defects, which depends on the state of structural disorder through $\theta_c$, the defect energy, governed by $\Lambda$ and $\B M$, and the reversible elastic energy, governed by $\B E^e$.

\section{Fourier-type relations}

\begin{figure}
\begin{center}
a) \epsfig{file=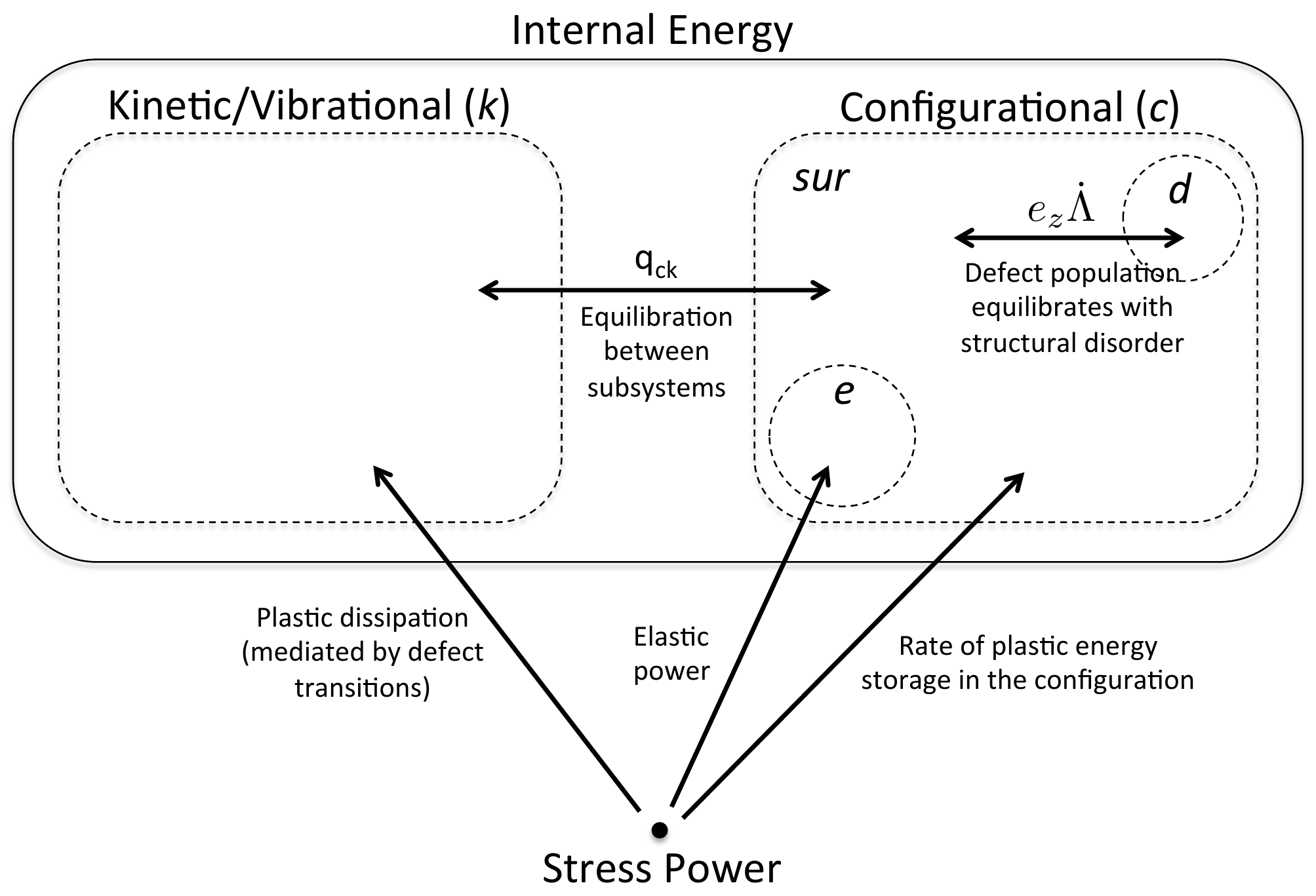, width=3.5in} \  b)  \epsfig{file=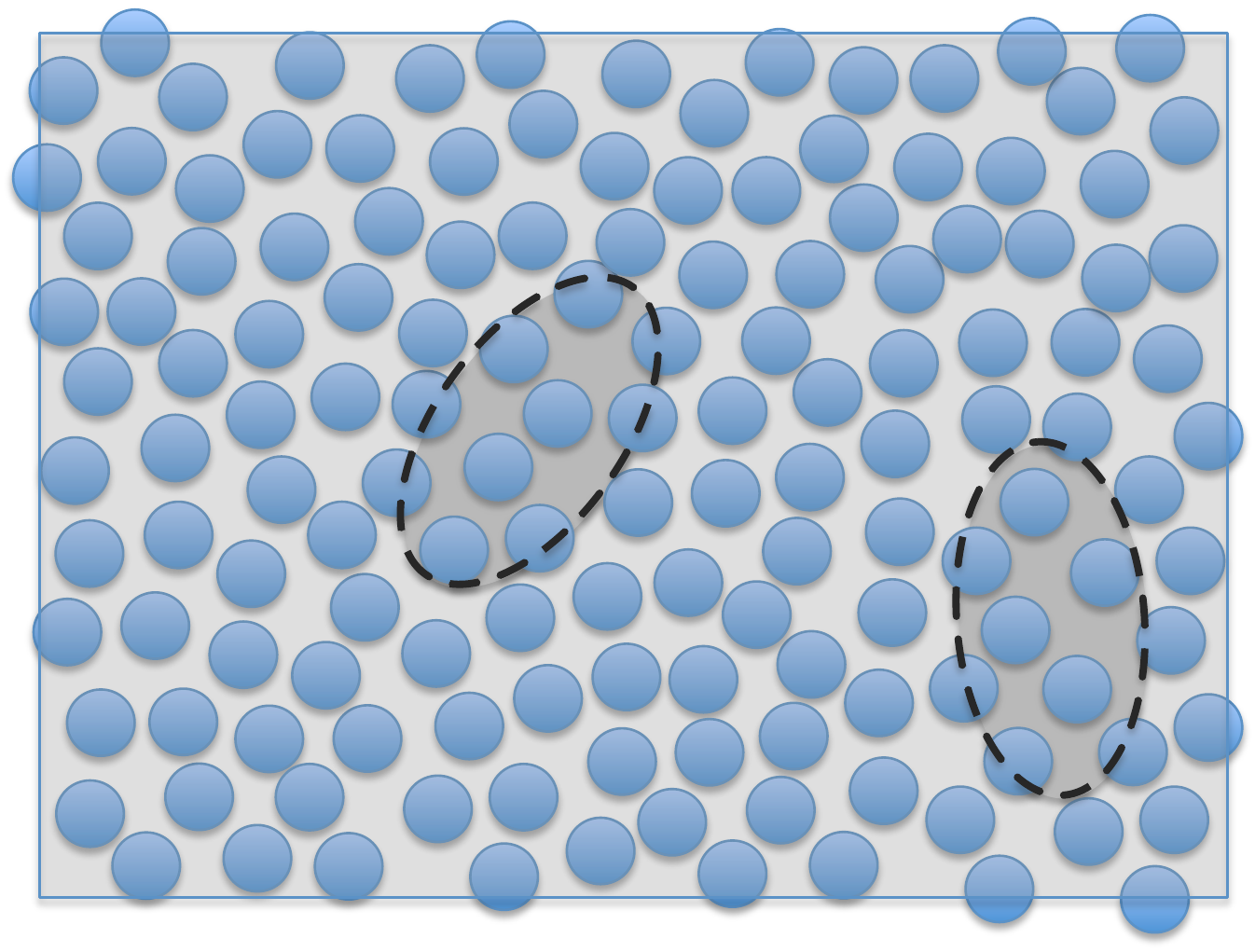, width=2.35in}
\caption{a) Energy transfer diagram, neglecting spatial fluxes, reflecting the transfer of energy from stress-working into the various internal energy mechanisms described in the current framework, as well as the exchange of power between these mechanisms. b) STZ's (illustrated schematically by the shaded ellipses) within a glassy material structure, treated as the flow defects, are produced/annihilated through equilibration with the configurational state of the surrounding material.} \label{Fig2}
\end{center}
\end{figure}

Our strategy to prevent a violation of the second law inequality in (\ref{second3}) is the same as the one discussed in Section \ref{sec:conventional}. We constrain the last four terms to be non-negative under all circumstances and guarantee non-negativity of the other terms together; it is a sufficient strategy for thermodynamic compatibility though not a unique one. Non-negativity of the second-row terms is guaranteed by adopting the ``Fourier-type relations'' in Eqs. (\ref{Fourier_conventional}), together with
\begin{align}
\dot\Lambda = \kappa_{cd}\left[\Lambda_0
e^{-\frac{e_z}{k_B \theta_c}} - \Lambda \right]\ ,& \ \ \ \
\kappa_{cd}=\hat{\kappa}_{cd}(\theta_c,\theta_k,d^p) \ge 0, \quad\quad
\Lambda_0=\hat{\Lambda}_0(m) = \frac{e^{(1-m^2) s_0}}{v_0} \ .\label{kcd}
\end{align}
Note that at fixed $m$, the condition for $\dot\Lambda\!=\!0$ is an equilibrium number of defects \emph{determined by the configurational temperature} through a Boltzmann-like factor $\sim\!e^{-\frac{e_z}{k_B \theta_c}}$. This statistical thermodynamics result, alluded to in Section \ref{sec:conventional}, goes beyond conventional modelling and is one of the most important consequences of the proposed framework.
The various modes of energy transfer in this model are visualized in Fig. \ref{Fig2}a.

\section{Dissipation inequalities and heat equations}

Our goal in this section is to complete the thermodynamic formulation by deriving dissipation inequalities and heat equations, similarly to the procedure discussed in Section \ref{sec:conventional}. To further exploit the second law inequality to derive additional constraints on the constitutive law, i.e. dissipation inequalities, we note that since all of the terms in the second row of the inequality in (\ref{second3}) were already made non-negative above, it is sufficient to satisfy the following reduced inequality
\begin{eqnarray}
\label{second4}
&&-2 k_B s_0\, \Lambda\, \B M:{\dot{\B M}}+\left(\frac{1}{\theta_c}\B T^p_{store}+\frac{1}{\theta_k}\B T^p_{dis}\right):\B D^p \ge 0 \ ,
\end{eqnarray}
where we have replaced $\B L^p$ with its symmetric part $\B D^p$ because it is contracted with symmetric tensors. To proceed, we need to discuss the evolution equation for $\B M$, which represents the average orientation of the flow defects population. We propose a minimal evolution rule of the form
\begin{equation}
\label{assumed_dotM}
\dot{\B M} = a \B D^p - b \B M \ ,
\end{equation}
where $a$ and $b$ are positive scalar functions. This equation is shown in Appendix \ref{STZ_app} to emerge on its own from a mesoscopic model of STZ dynamics.  Essentially, the first term produces orientational memory in accordance with the direction and rate of flow, and the latter is a relaxation term that pushes the system to lose orientational memory over time. Using the last equation with Eq. (\ref{etacderivs}), we obtain $\pa \hat\eta_c/\pa \B M:\dot{\B M}\! =\!-2k_Bs_0\,a\, \Lambda\,\B M:\B D^p + 2k_Bs_0\,b\,\Lambda\, m^2$. Substituting this expression into the inequality in (\ref{second4}) yields
\begin{eqnarray}
2k_Bs_0\,b\,\Lambda\, m^2 + \left(
-2k_Bs_0\,a\,\Lambda\, \B M+\frac{1}{\theta_c}{\B T}^p_{store}+\frac{1}{\theta_k}{\B T}^p_{dis}\right):\B D^p \ge 0.
\end{eqnarray}
The first term is always non-negative so it is sufficient to ensure the remaining term is non-negative
\begin{equation}
\label{plast_dissip}
\left(- 2k_Bs_0\,a\,\Lambda\,\B M +\frac{1}{\theta_c}{\B T}^p_{store}+\frac{1}{\theta_k}{\B T}^p_{dis}\right)\!:\!\B D^p \ge 0 \ .
\end{equation}
In the next section we will choose functional forms consistent with this inequality.

The last step in the thermodynamic development is the derivation of heat equations using the first law relations
in Eqs. (\ref{firstkI})-(\ref{firstcI}).
We focus first on Eq. (\ref{firstkI}).
Recalling that $\epsilon_k\!=\!\hat\epsilon_k(\theta_k)$ and taking into account the Fourier-type relations in Eq. (\ref{Fourier_conventional}), we obtain
\begin{equation}\label{heatk}
{c}_k \dot\theta_k =
{\B T}^p_{dis}: \B L^{p}
+{\kappa}_{ck}\ (\theta_c-\theta_k)+ \text{Div}_I\left[\kappa_k \,\B g_k\right] \ ,
\end{equation}
with $c_k\!=\!\hat{c}_k(\theta_k)\!\equiv\!\pa\hat\epsilon_k/\pa\theta_k$. This equation is essentially the same as Eq. (\ref{conventional_heat_k}), discussed in the context of the simple model in Section \ref{sec:conventional}. Hats are included to remind us which terms have been assigned functional dependences and which are independent variables (in the current stage of the analysis).  This is an ordinary heat equation for $\theta_k$ in which the plastic dissipation $\B T^p_{dis}\!:\!\B L^p$ and the heat flow between the kinetic/vibrational and configurational subsystems appear as source terms.

More interestingly,  for the configurational subsystem we apply the chain-rule to express $\dot{\epsilon}_c$ in terms of $\dot{\theta}_c$ and simplify the result using Eq. (\ref{firstcI}) to obtain
\begin{align}
c_c \dot\theta_c =&\B T^p_{store}:\!\B L^p
-{\kappa}_{ck}\ (\theta_c-\theta_k)-e_z  {\dot\Lambda} +
\text{Div}_I \left[{\kappa}_c\,  \B g_c
\right] \ ,\label{heatc}
\end{align}
with $c_c\!=\!\hat{c}_c(\theta_c,\B E^e,\Lambda,\B M)\!\equiv\!\pa\hat\epsilon_c/\pa\theta_c$. This is an evolution equation for the temperature of the configurational degrees of freedom, $\theta_c$, and hence describes the evolution of structural disorder. This is a central result within the present approach and is closely related to Eq. (\ref{conventional_heat_c}), discussed in the context of the simple model in Section \ref{sec:conventional}. This section concludes the derivation of the thermodynamic framework.

\section{An example: A constitutive model with internal variables}
\label{STZ_sec}
\noindent The constitutive model developed so far, along with the relevant balance laws, takes the form:\\

\noindent Conservation of linear and angular momentum: $\quad\quad\nabla\cdot\B T +\B b = \rho\dot{\B v}\ ,\  \ \ \B T = \B T^T$.\vspace{0.1cm}\newline

\noindent Kinematic decomposition and definition of intermediate space variables: $~~\partial \B \chi/\partial\B {X}\equiv\B {F}={\B F}^e\B {F}^p$, \newline $J^e\!=\!\det \B F^e, \ \ \ \B {C}^e\!=\!\B {F}^{eT}\B {F}^e, \ \ \ \B E^e\!=\!(\B {C}^e-\B {1})/2, \ \ \
 \B T^e\!=\!J^e\B {F}^{e-1}\B T\B {F}^{e-T}, \ \ \ \B T^p\!=\!\B {C}^e\B T^e, \ \ \ \B L^p\!=\!\dot{\B {F}}^p\B {F}^{p-1}$. \vspace{0.1cm}\newline

\noindent Elasticity relation: \quad\quad\quad\quad\quad\quad\quad~$\quad\quad\B T^e =2 G \B E^{e \prime}+\kappa (\text{tr}\B E^e)\mathbf{1}$. \vspace{0.1cm}\newline

\noindent Decomposition of the plastic stress: $\quad\quad\B T^p\!=\!\B T^p_{store}+\B T^p_{dis}$. \vspace{0.1cm}\newline

\noindent Evolution law for the defect concentration internal variable: $\dot\Lambda\!=\!\kappa_{cd}\left[\Lambda_0(m)e^{-\frac{e_z}{k_B \theta_c}}\!-\!\Lambda \right]$ with $\Lambda_0(m)\!=\!\frac{e^{(1-m^2) s_0}}{v_0}$. \vspace{0.1cm}\newline

\noindent Kinetic/vibrational heat equation: $\quad\quad{c}_k \dot\theta_k \!=\! {\B T}^p_{dis}: \B L^{p}+{\kappa}_{ck}\ (\theta_c-\theta_k)+ \text{Div}_I\left[\kappa_k \,\B g_k\right]$. \vspace{0.1cm}\newline

\noindent Configurational heat equation: $\quad\quad\quad~~{c}_c \dot\theta_c\!=\!\ \B T^p_{store}:\!\B L^p -\kappa_{ck}\ (\theta_c-\theta_k)-e_z  \dot\Lambda + \text{Div}_I \left[\kappa_c\,  \B g_c \right]$.\\

The  not yet determined functions for the conduction coefficients $\kappa_{ck}$, $\kappa_{cd}$, $\kappa_c$, $\kappa_k$ must all be non-negative. The functions for $\B D^p$ and $\B T^p_{store}$, and the coefficients in the $\dot{\B M}$ relation, have to be derived from physical considerations and shown to satisfy the constraint of Eq. (\ref{plast_dissip}). While our focus in the present paper is on the thermomechanical framework, and by no means on specific constitutive models, we aim here at discussing a more elaborated internal variables model that goes beyond the simple example discussed in Section \ref{sec:conventional}. Below we present a constitutive model based on an STZ micro-mechanism, which rather naturally fits in the thermomechanical framework. The resulting complete set of equations will allow us in the next section to present a few illustrative applications of the model (full details of the derivation can be found in Appendix \ref{STZ_app}).

Based on extensive observations, we state that irreversible flow in amorphous solids is mediated by a population of flow defects in the form of oriented STZ's --
sparsely distributed localized zones that are more susceptible to shear-driven transformations in a particular direction than their surrounding material (see Fig. \ref{Fig2}b). We view STZ's not as permanent flow defects, but rather as configurational fluctuations that can appear and disappear during flow and/or in response to thermal fluctuations. We adopt the main physical ingredients of the STZ theory, which has been built on the flow-defect theories of Turnbull, Cohen, Argon, Spaepen and others \citep{Turnbull1970, Spaepen1977, Argon1979, ArgonShi1983}, and incorporate them into the presently developed framework. We define the microscopic parameters: $v_0$ is the volume of an STZ (consistent with its prior usage), $e_z$ is the characteristic energy required to produce an STZ, $\Delta$ is the energy required to shear-transition an STZ along its weak direction, $\gamma_s$ is the half-amplitude of the shear strain of an STZ transition, and $\nu$ is an attempt frequency for STZ transitions. One version of the two-dimensional STZ model leads the following equations (see Appendix \ref{STZ_app}):\\

\noindent  Evolution of the orientation internal variable:
\begin{equation}
\label{mevolve1}
\dot{\B M}= \frac{1}{\Lambda} \left(\frac{\B D^p}{v_0\, \gamma_s} -\kappa_{cd} \Lambda_0
e^{-\frac{e_z}{k_B \theta_c}} \B M \right) \ ,
\end{equation}
which is consistent with Eq. (\ref{assumed_dotM}) and allows a unique determination of $a$ and $b$ there.\\

\noindent  Flow rule:
\begin{equation}
\label{D_1st1} \B D^p = \frac{2 v_0\gamma_s \nu}{\tau_0}\,\Lambda\, \,
e^{-\Delta/k_B\theta_k}\left(\B T^{p\ \prime}_{dis}
-\tau_0 \B M\right) \ \ \ \text{with} \ \ \ \tau_0=2k_B\theta_k/v_0\gamma_s.
\end{equation}
These two constitutive equations are then substituted into the second law inequality of (\ref{plast_dissip}), which can be satisfied by choosing the following form for $\B T^p_{store}$.\\

\noindent Stored plastic stress:
\begin{equation}
\label{Tp_store_result1}
 \B T^p_{store}=\frac{2 k_B \theta_c}{v_0\, \gamma_s} (s_0-1)\ \B M \quad\hbox{with}\quad s_0 > 1 \ .
\end{equation}

\section{Basic properties of the internal variables constitutive model}

Our goal here is to discuss the properties of the internal variables constitutive model described above, focussing on features that go beyond those already discussed in the context of the simple model in Section \ref{sec:conventional}. In particular, the discussion of thermodynamic aging and configurational temperature diffusion applies here as is, since these are generic features of the proposed framework. The existence of a steady state configurational temperature is also generic, though we discuss it here again in order to gain a somewhat better physical understanding of its emergence.

\subsection{Vicinity of the glass transition: Newtonian viscosity}
\label{viscosity}

Near the glass transition temperature, relaxation times are large (macroscopic), but still fall within the experimental measurement window and phenomena such as Newtonian flow and aging can be probed \citep{Angell1988, Gotze1992, Angell2000, Lu2003, Cavagna2009}. Our goal here is to explore the emergence of such phenomena within the proposed constitutive law.  We first focus on Newtonian flows and hence consider spatially homogeneous, pure shear situations.  We reduce the analysis to two dimensions of space and assume the deformation is driven by a deviatoric stress of the form $\B T\!=\!T\ \B d $, where $d_{ij}\!=\!(2e_ie_j-\delta_{ij})/\sqrt{2}$ is the unit-magnitude deviatoric tensor with major principal direction $\B e$.

Then we can write $\B D^p\!=\!D^p\ \B d$, $\B M\!=\!M\ \B d$, and $\B T^p_{\{\cdot\}}\!=\! T^p_{\{\cdot\}}\B d$ for ``$\cdot$'' indicating any of our subscripts. The viscoplastic relation in Eqs. (\ref{mevolve1})-(\ref{D_1st1}) gives
\begin{eqnarray}
\label{simple_shear}
\dot{M} \!=\! \left[\frac{D^p}{\Lambda_s v_0\gamma_s} - \kappa_{cd}(\theta_c,\theta_k,d^p)\, M \right],\quad
D^p \!=\! 2\,v_0\gamma_s\,\Lambda_s\, \nu\, e^{-\Delta/k_B\theta_k}\left[\frac{T^p_{dis}}{\tau_0}-M\right] \ ,
\end{eqnarray}
where we have assumed Eq. (\ref{kcd}) remains at its fixed-point $\Lambda\!=\!\Lambda_0 e^{-\frac{e_z}{k_B \theta_c}}\equiv\Lambda_s$. Also, recall that $d^p\!=\!|D^p|$.

A crucial step in analyzing these equations is specifying the thermodynamic coupling coefficient $\hat{\kappa}_{cd}(\theta_c,\theta_k,d^p)$ that is responsible for the creation and annihilation of STZ's. As was discussed in detail above, this (and other) thermodynamic coupling coefficient depends explicitly on the plastic rate of deformation $d^p$. Moreover, in the absence of plastic deformation, $d^p\!=\!0$, $\kappa_{cd}$ may be finite due to thermal fluctuations alone. Therefore, we set
\begin{equation}
\label{k_cd}
\hat{\kappa}_{cd}(\theta_c,\theta_k,d^p) = d^p \hat{f}(\theta_c) + \nu\,\hat{\zeta}(\theta_k) \ ,
\end{equation}
where the dimensionless function $\hat{f}(\theta_c)$ remains an unspecified function of the configurational temperature $\theta_c$. It is reasonable that the thermal attempt frequency $\nu$ sets the time-scale for the $\theta_k$-dependent term, and the flow-rate plays the analogous role for the $\theta_c$-dependent term.
The physical meaning of the dimensionless function $\hat{\zeta}(\theta_k)$ will be discussed soon. We are now in a position to study Eqs. (\ref{simple_shear}) in various situations. First, consider the steady state linear response -- i.e. small variations about $\{M\!=\!0, d^p\!=\!0, \Lambda\!=\!\Lambda_s, \theta_c\!=\!\theta_k\}$ -- to a small applied stress $T^p$. The Newtonian (extensional) viscosity is defined by $\mu_{_N} \!\equiv\! \lim_{d^p\to 0} T^p/D^p$.

To calculate $\mu_{_N}$, we first substitute Eq. (\ref{k_cd}) in Eq. (\ref{simple_shear}) to obtain
\begin{equation}
\dot{M} = \left[\frac{D^p}{\Lambda v_0\gamma_s} - d^p\,M\,\hat{f}(\theta_c) - \nu\,\hat{\zeta}(\theta_k)\,M \right] \ .
\end{equation}
As we are interested in the linear response regime, the  second-order term proportional to $d^p\,M$ can be omitted. Moreover, as we focus on steady state we can set $\dot{M}\!=\!0$ and use $D^p$ of Eq. (\ref{simple_shear}) to obtain
\begin{equation}
\label{m_eta}
M = \frac{T^p_{dis}}{\tau_0} \frac{1}{1 + \frac{1}{2}\hat{\zeta}(\theta_k)\,e^{\Delta/k_B\theta_k}}  \ .
\end{equation}
In the vicinity of the glass transition temperature and just below it, we expect spontaneous thermally-activated creation and annihilation of STZ's to be strongly suppressed, i.e. $\hat{\zeta}(\theta_k)$ becomes small. Therefore,  $\hat{\zeta}(\theta_k)\,e^{\Delta/k_B\theta_k}\!\ll\!1$ is a small parameter and to leading order in the smallness Eq. (\ref{m_eta}) reads
\begin{equation}
\frac{T^p_{dis}}{\tau_0} - M \simeq  \frac{T^p_{dis}\,\hat{\zeta}(\theta_k)\,e^{\Delta/k_B\theta_k}}{2\tau_0} \ .
\end{equation}

By comparing the left side to the expression for $D^p$ in Eq. (\ref{simple_shear}), we immediately obtain $D^p\!\propto\!\hat{\zeta}(\theta_k)$, which tells us that the material flows persistently only because $\hat{\zeta}(\theta_k)$ is finite. The physics here is clear: when $\hat{\zeta}(\theta_k)$ vanishes, STZ's cannot be created and annihilated by thermal fluctuations (recall that mechanically-induced STZ creation and annihilation do not contribute at linear order) and hence existing (properly oriented) STZ's can undergo transitions in the direction of the applied stress, but once they all do so, plastic flow cannot persist. To complete the calculation of $\mu_{_N}$ we use Eq. (\ref{Tp_store_result1}) for $T^p_{store}$, which together with Eq. (\ref{m_eta}) implies $T^p_{store}\!\approx\! const. \times T^p_{dis}$ and hence $T^p\!=\!T^p_{store}\!+\!T^p_{dis}\!\approx\! const. \times T^p_{dis}$. Therefore, we obtain
\begin{equation}
\label{STZ_eta}
\mu_{_N} \approx \frac{T^p_{dis}+T^p_{store}}{2\,v_0\gamma_s\,\Lambda_s\, \nu\, e^{-\Delta/k_B\theta_k} \left(\frac{T^p_{dis}\,\hat{\zeta}(\theta_k)\,e^{\Delta/k_B\theta_k}}{2\tau_0}\right)} \propto \frac{\tau_0\,e^{\frac{e_z}{k_B \theta_k}}}{\nu\,\hat{\zeta}(\theta_k)} \ ,
\end{equation}
where we used $\theta_c\!\simeq\!\theta_k$ in the linear response regime and the definition of $\Lambda_s$.
We observe that $\mu_{_N}\!\sim\!\hat{\zeta}(\theta_k)^{-1}$, which shows that the Newtonian viscosity increases as $\hat{\zeta}(\theta_k)$ decreases with decreasing temperature $\theta_k$ and strictly diverges if $\hat{\zeta}(\theta_k)$ vanishes.
Indeed, the essence of the glass transition puzzle, which is commonly probed by measuring the dramatic increase of $\mu_{_N}$ with decreasing temperature $\theta_k$, is encapsulated in $\hat{\zeta}(\theta_k)$, which at present cannot be calculated from first principles. In the liquid state (at high temperatures) $\hat{\zeta}(\theta_k)\!\simeq\!1$, while $\hat{\zeta}(\theta_k)$ decreases dramatically in the supercooled and glassy regime (for ``fragile'' glasses \citep{Angell1988}). We also note that $\mu_{_N}$ strictly diverges only if there exists an ideal glass transition, $\hat{\zeta}(\theta_k)\!=\!0$, and that due to the appearance of the Arrhenius factor $e^{\Delta/k_B\theta_k}$ in Eq. (\ref{STZ_eta}), $\hat{\zeta}(\theta_k)$ can be identified as the non-Arrhenius temperature dependence of the Newtonian viscosity, similarly to \citet{Langer2008}.

\subsection{Low temperatures: The emergence of a flow stress for steady shearing}
\label{lowT}
\begin{figure}
\begin{center}
\epsfig{file=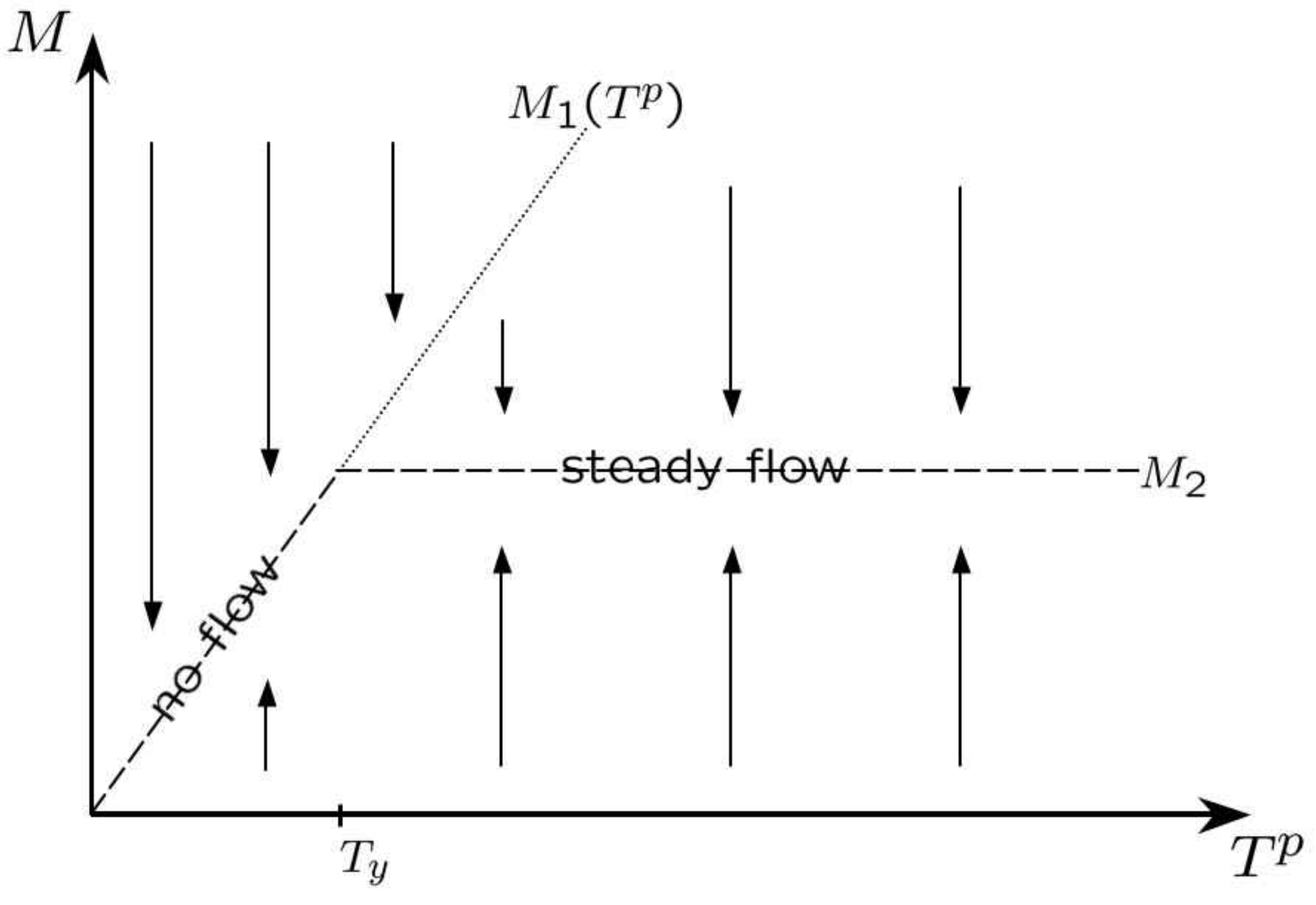, width=3.5in}
\caption{Stability diagram indicating the dynamics of $M$ in Eq. (\ref{stability_exchange}) for any applied stress $T^p$. Dashed lines represent stable fixed points, the dotted line represents (marginally) unstable fixed points.}\label{Fig3}
\end{center}
\end{figure}

We would like now to explore the nonlinear response in the framework of the proposed model, and in particular the emergence of a yield stress. Using Eqs. (\ref{Tp_store_result1})-(\ref{simple_shear}), we obtain
\begin{equation}
\label{dp_yielding}
D^p \propto T^p - \left[\tau_0+2(s_0-1)k_B \theta_c/v_0 \gamma_s\right]M \equiv T^p - \bar\tau M \ .
\end{equation}
We focus on low bath temperatures $\theta_k$ for which $\hat{\zeta}(\theta_k)$ is extremely small and can be neglected. In this case, Eq. (\ref{k_cd}) can be approximated as $\hat{\kappa}_{cd}(\theta_c,\theta_k,d^p)\!\simeq\!d^p \hat{f}(\theta_c)$ and the first equation in (\ref{simple_shear}) takes the form
\begin{equation}
\label{stability_exchange}
\dot{M} \propto \left[T^p - \bar\tau M\right]\left[\left(\Lambda\,v_0\gamma_s\,\hat{f}(\theta_c) \right)^{-1} - M\  \text{sign}\left(T^p - \bar\tau M\right) \right]\ ,
\end{equation}
where $d^p\!=\!D^p\ \text{sign}(D^p)$ and Eq. (\ref{dp_yielding}) were used.

To understand how a flow resistance naturally emerges within the model in the nonlinear response regime, we need to analyze the fixed-points of Eq. (\ref{stability_exchange}) and their stability (see Fig. \ref{Fig3}). The first bracketed term in Eq. (\ref{stability_exchange}) vanishes only at the fixed point $M_1\!=\!T^p/\bar\tau$.
In the second bracketed term, $\Lambda$ reaches its fixed-point at $\Lambda_s\!=\!\hat{\Lambda}_0(m)\,  e^{-\frac{e_z}{k_B \theta_c}}$, which depends itself on $M$. The $M$-dependence of $\Lambda_0$ is Gaussian (cf. Eq. (\ref{kcd})), so we expect the second bracketed term to have zero or two positive solutions. {{We assume that the yet to be determined function $\hat{f}(\theta_c)$ is such that only one solution exists in the physical range $|M|\!<\!1$. While this condition is not necessary on general grounds, it will be shown below to give rise to a physically appealing picture of yielding. This single physical solution is denoted as $M\!=\!M_2$ and satisfies
\begin{equation}
\frac{M_2 \, \hat{\Lambda}_0(|M_2|)}{\text{sign}\left(T^p - \bar\tau M_2\right) }=  \frac{e^{\frac{e_z}{k_B \theta_c}}}{v_0\gamma_s\,f(\theta_c)} \ .
\end{equation}
While $\theta_c$ is indeed a dynamic variable in the above expression, it is expected to vary slowly compared to $M$, so we treat it as a constant for the ensuing stability analysis.}}

For stresses $T^p$ such that $M_1\!<\!M_2$, $M_1$ is the only stable fixed-point, see Fig. \ref{Fig3}. Then Eq. (\ref{dp_yielding}) implies the system approaches $D^p\!=\!0$. This is the sub-yielding regime in which after transient plastic deformation the flow arrests. As the applied stress $T^p$ further increases, we reach the point where $M_1\!=\!M_2$. At this point an exchange of stability occurs, and $M_2$ becomes the stable fixed-point, see Fig. \ref{Fig3}. In this case, Eq. (\ref{dp_yielding}) shows that $D^p\!>\!0$, i.e. that the material flows persistently. This happens at a threshold stress $T^p\!=\!T_y$ for
\begin{equation}
\label{yield_stress}
T_y = \bar\tau M_2 \ ,
\end{equation}
which we interpret as a dynamic yield stress, responsible for the transition from non-flowing to flowing steady states. It is apparently a function of the state of disorder through $\theta_c$, ultimately through the details of $\hat{f}(\theta_c)$, which cannot be determined through thermodynamics alone. It is clear, that Eqs. (\ref{stability_exchange})-(\ref{yield_stress}) make physical sense if $\hat{f}(\theta_c)$ leads to a solution with $|M_2|<1$. In summary, we have shown that a steady-state flow resistance in the model emerges as a dynamic exchange of stability in the STZ orientation variable, and need not be stipulated a priori as done, for example, in the simple model discussed in Section \ref{sec:conventional}.

\subsection{Rejuvenation and the steady state of configurational disorder}
\label{rejuvenation}

The steady state of configurational disorder, i.e. of $\theta_c$, was discussed in Section \ref{sec:conventional} in the context of the simple phenomenological model, demonstrating its generic features. We briefly discuss this issue again here in the context of a the internal variables model. To that aim, we focus again on the configurational heat equation in (\ref{heatc}) and observe that under rather generic situations in which $\dot\Lambda\!>\!0$, $\B M\!:\!\dot{\B M}\!>\!0$ and $\theta_c\!>\!\theta_k$, the only potentially positive term that can be responsible for rejuvenation is $\B T^p_{store}\!:\!\B L^p$. When rejuvenation takes place, it cannot persist forever and $\theta_c$ reaches a steady state under persistent plastic deformation, as is observed in numerous computer simulations of glass forming systems, e.g. those of \citet{Ono2002, HaxtonLiu2007}. To see how it emerges, we specialize to pure shear isothermal (in $\theta_k$) situations as above, which allows us to omit all time and space derivatives in Eq. (\ref{heatc}) to obtain
\begin{eqnarray}
0 = \tau^p_{store}d^p-\kappa_{ck}(\theta_c-\theta_k) \ . \label{heatc_SS}
\end{eqnarray}
$\B L^p$ was replaced by its symmetric part $\B D^p$ because it is contracted with a symmetric tensor and we have simplified $T^p_{store}D^p$ to $\tau^p_{store}d^p$, since steady pure shearing under the prescribed driving guarantees positivity of $T^p_{store}$ and $D^p$. 

To actually calculate the steady-state value, we recall that (as discussed several times above) the thermodynamic coupling coefficient $\kappa_{ck}$ includes a deformation-driven part that depends on $d^p$ and a thermally-driven part that depends on $\theta_k$. At sufficiently small $\theta_k$, we have
$\kappa_{ck}\!\simeq\!d^p\,k_B\,\hat{h}(\theta_c)/v_0\gamma_s$, where $\hat{h}(\theta_c)$ is a positive dimensionless function. This leads to
\begin{equation}
\theta_c = \theta_k + \frac{\tau^p_{store}}{\hat{h}(\theta_c)} = \theta_k + \frac{2(s_0-1)\theta_c |\hat{M}_2(\theta_c)|}{\hat{h}(\theta_c)} \ ,
\end{equation}
which is an implicit equation for the steady state value  of $\theta_c$. The most interesting aspect of this result, beyond the existence of a steady state (which is expected on general grounds), is that the steady state value of $\theta_c$ is independent of the plastic deformation rate $d^p$ and the externally applied stress. For sufficiently low temperatures, $\theta_k$ can be also neglected. These results are consistent with many numerical simulations as long as $1/d^p$ is much larger than molecular timescales, see for example \cite{Ono2002}.

\section{Summary and discussion}
\label{summary}

In this paper we have developed systematically a finite-deformation, nonequilibrium, continuum thermomechanical framework for deforming solids wherein the configurational degrees of freedom evolve slowly compared to kinetic/vibrational degrees, and may be characterized by a temperature of their own. This description of a solid as composed of weakly interacting configurational and kinetic/vibrational subsystems is inspired by recent progress in statistical physics \citep{Teff_review_2011}. Within the proposed framework, internal variables play an essential role in characterizing the out-of-equilibrium state of a deforming solid and are explicitly associated with distinct energy and entropy contributions. An important deviation from standard thermodynamics is that coupling and transport coefficients associated with the configurational subsystem feature an explicit dependence on the plastic rate of deformation.

We derived a configurational heat equation, in addition to an ordinary heat equation, and showed how a strain-rate independent steady state of the configurational temperature naturally emerges under persistent deformation.
First, this general framework was applied to a simple phenomenological two-temperature plasticity model. Then, building on the concept of Shear-Transformation-Zones (STZ's), and invoking a minimal scenario involving only one scalar and one tensorial internal variables, an internal variables based constitutive law was developed. These constitutive models were shown to capture various amorphous plasticity phenomena, including isotropic and kinematic strain softening/hardening, Newtonian viscosity and thermodynamic aging near the glass temperature, a sharp yield stress at low temperatures, flow driven structural rejuvenation and memory effects. In addition, as was discussed in Section \ref{sec:conventional}, the diffusion of the configurational temperature -- with the accompanying configurational heat conductivity $\kappa_c$ -- might offer a new route to describing size effects in constitutive modelling.

We hope that the present development can serve as a physically-motivated and self-consistent thermomechanical framework for formulating constitutive models for the elasto-viscoplastic deformation of solids that lack internal equilibrium. This should eventually allow to address a wide range of physical phenomena related to the shape deformations and failure modes of real amorphous materials (e.g. bulk metallic glasses), which find an enormous range of modern engineering applications, but whose faithful description in the context of complex realistic problems is currently missing. In order to make progress in achieving this ambitious goal, we plan to follow up the present investigation in several directions. First, as was mentioned above, we plan to address pressure effects and the coupling between shear and volumetric deformation, which were not discussed here. This is an important aspect of the deformation of amorphous solids and hence should be systematically incorporated into the proposed thermomechanical framework and constitutive modeling.

In addition, we need to perform a quantitative analysis of available experimental data to determine specific material parameters and functions, and to test the extent to which the proposed model can quantitatively account for various experimental protocols (quasi-static loading, dynamic loading, strain-rate controlled experiment, stress controlled experiment, jumps in the applied strain-rate, creep, strain relaxation) and control parameters (strain rate, deformation temperature, thermal history etc.). We then plan to consider spatially inhomogeneous deformation situations in 2D and to specifically focus on shear band widening dynamics, which involve a lengthscale associated with the diffusion of the configurational temperature. Finally, we plan to implement the proposed model in a Finite Element Method code to allow testing and parameters calibration in various 3D problems. We hope to be able to report on some progress along these lines in the not too far future.

\newpage

\noindent{\bf Acknowledgements}\\

K.K. acknowledges funding from NSF CBET grant 1253228. E.B. acknowledges support from the Minerva Foundation with funding from the Federal German Ministry for Education and Research, the Israel Science Foundation (Grant No. 712/12), the Harold Perlman Family Foundation and the William Z. and Eda Bess Novick Young Scientist Fund.

\vspace{0.5cm}

\appendix
\numberwithin{equation}{section}

\noindent{\bf Appendix}
\section{STZ-based flow rule}\label{STZ_app}

Our goal in this section is to augment our thermomechanical model with ideas from recent Shear-Transformation-Zone (STZ) theories \citep{Falk1998, Falk2011, Pechenik2003, Langer2004, Pechenik2004, Pechenik2005, STZ_CircularHole_2007, STZ_ShearBanding_2007, BLPI2007, BLPII2007, STZ_cavitation_2008, HoleStability2008, Langer2008, Bouchbinder_Teff_2008, Daub2008, STZ_ShearBanding_2009, BLIII2009, Daub2009, Kovacs_Teff_2010, LinearResponse_PRL_2011, LinearResponse_PRE_2011, Rycroft2012}. By doing this, we provide additional information to fill-in remaining details of the constitutive relations, culminating in a shear flow constitutive law.  To motivate the connection, we first discuss the STZ micromechanism and its ability to replicate the $\dot{\Lambda}$ and $\dot{\B M}$ functional forms deduced in the previous sections.

We start by postulating the existence of a population of oriented STZ's --
sparsely distributed localized zones that are more susceptible to shear-driven transformations in a given direction than their surrounding material (see Fig. \ref{Fig2}b). We view STZ's not as permanent flow defects, but rather as configurational fluctuations that can appear and disappear due to thermal and mechanically-generated noise (i.e. the mesoscale stress fields that are generated by structural rearrangements). Therefore, the density of STZ's essentially describes the probability to find such a configurational fluctuation. The existence of STZ's has been demonstrated in numerous computer simulations and experimental systems, see for example \citet{ArgonQuo1979, Srolovitz1981, Srolovitz1983, Falk1998, Mayr2006, Schall2007, STZ_BMG_2008, Delogu2008}. Below we will adopt the main physical ingredients of the STZ theory, which has been built on the flow-defect theories of Turnbull, Cohen, Argon, Spaepen and others \citep{Turnbull1970, Spaepen1977, Argon1979, ArgonShi1983}, and incorporate them into the presently developed framework.

In two dimensions, each STZ -- during its lifetime -- is described by its orientation $\phi$ with respect to a fixed
reference direction. By definition, an STZ oriented along $\phi$ is
predisposed to undergo a shear (volume-conserving) transformation in the direction of the purely
deviatoric, symmetric, unit-magnitude tensor $\B d(\phi)$ whose major principal
vector $\B e(\phi)$ is oriented along the $\phi$
direction. The components of the tensor take the form $d_{ij}(\phi)\!=\!(2 e_i(\phi)e_j(\phi)\!-\!\delta_{ij})/\sqrt{2}$. The population density of STZ's at position $\B x$ at time $t$ oriented along $\phi$ is defined as $n(\phi; \B x, t)$, but we will generally write this without
the $\B x$ and $t$ for notational simplicity. Note that azimuthal periodicity implies $n(\phi)\!=\!n(\phi+\pi)$.

We define $\Lambda$ to be the (intermediate) volume density of STZ's, expressed as an integral over all orientations
\begin{equation}\label{Lambdadef}
\Lambda=\int_{-\pi/2}^{\pi/2}n(\phi)\ d\phi \ .
\end{equation}
Similarly, we define $\B M$ to be a normalized tensorial moment of
the STZ distribution in the sense that
\begin{equation}\label{Mdef}
\B M=-\frac{1}{\Lambda}\int_{-\pi/2}^{\pi/2} n(\phi)\ \B
d(\phi) \ d\phi \ ,
\end{equation}
where the minus sign is used for consistency with previous analyses \citep{Falk2011}. Note also that by construction $|\B M|\!\le\!1$ and that $\B M$ bears some resemblance to orientational order parameters in magnetic and liquid crystalline systems.

The STZ's are assumed to be two-state systems that can be driven to ``flip'' from their original orientation $\phi$ to a perpendicular orientation $\phi+\pi/2$ during their lifetime. While this is definitely an oversimplification as the STZ's are multi-atom objects that can feature nontrivial internal structures, this assumption was shown to be sufficient to capture a range of glassy deformation phenomena \citep{Falk2011}. One might characterize an STZ  oriented along $\phi$ as an elliptical region whose semi-major axis aligns with $\B e(\phi)$. A ``flip'' is a pure shear transition, that deforms the STZ inward along the  $\B e(\phi)$ direction and outward along  $\B e(\phi+\pi/2)$, leaving the STZ, post-transition, oriented in the $\B e(\phi+\pi/2)$ direction.  The rate at which STZ's oriented along $\phi$ undergo a shear transformation into a state oriented along $\phi\!+\!\pi/2$ depends on the resolved stress, $\B T^p_{dis}\!:\!\B d(\phi)$, and thermal fluctuations, as specified by a rate-factor $\C R (\B T^p_{dis}\!:\!\B d(\phi),\theta_k, \theta_c)$. For simplicity
we omit the arguments $\theta_k$ and $\theta_c$ in writing the rate-factor hereafter. Note that we assume that STZ transitions are driven by the dissipative stress $\B T^p_{dis}$. The evolution law for $n(\phi)$ must account for the change in the STZ's orientation when it undergoes shear transformations, as well as for
the creation and annihilation of STZ's due thermal and mechanically-generated structural fluctuations. Therefore, we suggest that the evolution of $n(\phi)$ satisfies the following master equation
\begin{equation}
\dot{n}(\phi)=\C R (\B T^p_{dis}:\B d(\phi+\pi/2))\
n(\phi+\pi/2)-\C R (\B T^p_{dis}:\B d(\phi))\ n(\phi) +
\kappa_{cd}\ \left(\frac{\Lambda_0}{\pi}e^{-\frac{e_z}{k_B
\theta_c}}-n(\phi)\right)  .
\end{equation}
Term by term, this equation represents the rate-of-change of $n(\phi)$ in terms
of the rate of STZ's transitions \emph{into} a state $\phi$, minus those
transitions \emph{out of} a state $\phi$, plus a production/annihiation term that tends to equilibrate the STZ's
population with $\theta_c$ in an isotropic manner. $\kappa_{cd}$ is a thermodynamic coupling coefficient that plays the role of a noise source that drives the creation and annihilation of STZ's. We must emphasize that
this form is consistent with the previous thermomechanical analysis; it agrees
exactly with our second-law constraint, Eq. (\ref{kcd}), i.e.
\begin{equation}
\label{Levolve}
\dot{\Lambda}=\int_{-\pi/2}^{\pi/2}\dot{n}(\phi)\
d\phi=\int_{-\pi/2}^{\pi/2} \kappa_{cd}\
\left(\frac{\Lambda_0}{\pi}e^{-\frac{e_z}{k_B
\theta_c}}-n(\phi)\right)\ d\phi= \kappa_{cd}\
\left(\Lambda_0e^{-\frac{e_z}{k_B \theta_c}}-\Lambda\right) \ .
\end{equation}
The integral simplifies as above because the transition terms do not affect the total number of STZ's

Macroscopic plastic deformation arises in this model from the accumulated effect of
all STZ's transitions  occurring at a given time. Hence, we express the plastic shear (deviatoric)
flow rate $\B D^p$ as
\begin{equation}
\label{int_flow}
\B D^p = 2\,v_0 \, \gamma_s \int_{-\pi/2}^{\pi/2}\C \  \C R(\B
T^p_{dis}:\B d(\phi))\ n(\phi)\  \B d(\phi)\  d\phi \ ,
\end{equation}
where $\gamma_s$ is (half) the shear strain magnitude of an STZ transition.

Taking the time derivative of $\B M$ defined in Eq. (\ref{Mdef}) and using Eqs. (\ref{Lambdadef})-(\ref{int_flow}), we obtain the following closed-form equation for $\dot{\B M}$
\begin{equation}
\label{mevolve}
\dot{\B M}= \frac{1}{\Lambda} \left(\frac{\B D^p}{\gamma_s\, v_0} -\kappa_{cd} \Lambda_0
e^{-\frac{e_z}{k_B \theta_c}} \B M \right) \ .
\end{equation}
It takes the form assumed in Eq. (\ref{assumed_dotM}), allowing us to identify
\begin{equation}
\label{a_and_b}
a = \frac{1}{\Lambda\,v_0\, \gamma_s} \quad\hbox{and}\quad b = \frac{\kappa_{cd} \Lambda_0
e^{-\frac{e_z}{k_B \theta_c}}}{\Lambda} \ .
\end{equation}

Equation (\ref{int_flow}) becomes an explicit flow rule when the function $\C R$ is specified. In the most general case this would lead to an integral representation of the shear flow rate, that is not necessarily reducible to a simple closed form relation in terms of $\Lambda$ and $\B M$ alone. In order to explore the properties of the emerging constitutive law, we proceed by choosing a prototypical form for the $n(\phi)$ distribution compatible with some
given $\Lambda$ and $\B M$ values:
\begin{equation}
\label{n_approx}
n(\phi)=\frac{\Lambda}{\pi}\left(1-2\, \B M :\B d(\phi)\right) \ .
\end{equation}

It is important at this point to reflect on the status of our analysis so far. We have at hand a flow rule (Eq. (\ref{int_flow}), together with Eqs. (\ref{Levolve}), (\ref{mevolve}) and (\ref{n_approx})),
which is not fully determined until the rate factor $\C R$ is specified. In addition, we still have to satisfy the second law inequality in (\ref{plast_dissip}), where ${\B T}^p_{store}$ is not yet specified
(note that ${\B T}^p_{dis}$ is known, once ${\B T}^p_{store}$ is specified, through the relation ${\B T}^p\!=\!{\B T}^p_{store}+{\B T}^p_{dis}$).
There are (at least) two possible routes to proceed; either we model ${\B T}^p_{store}$ based on physical,but not purely thermodynamical, considerations and use the second law inequality in (\ref{plast_dissip}) to constrain the rate factor $\C R$ or
we model the rate factor $\C R$ based on physical considerations and use the second law inequality in (\ref{plast_dissip}) to constrain ${\B T}^p_{store}$.
The two routes are expected to have different consequences and it is not a priori clear which one is more physically faithful.

Below we follow the second route by adopting a simple rate-factor $\C R$ and exploring the consequences through several approximations.  We consider a simple thermally activated process for STZ transitions, similar to the form for atomic jumps in \cite{Spaepen1977}, for which
\begin{equation}
\label{rate-factor}
\C R(\B T^p_{dis}:\B d(\phi),\ \theta_k)
=\nu\, \exp\left[-\frac{\Delta -v_0\, \gamma_s\ \B T^p_{dis}:\B
d(\phi)}{k_B\,\theta_k}\right] \ ,
\end{equation}
where $\nu$ is a molecular attempt rate and $\Delta$ is an STZ transition barrier.
Using this simple rate factor in the flow rule in Eq. (\ref{int_flow}), one can perform the integration analytically to obtain the following closed-form expression
\begin{equation}
\label{simple_flow}
\B D^p\!=\!2\, v_0\, \gamma_s\, \Lambda\,\nu\,e^{-\Delta/k_B\theta_k}\left[I_1\left(\frac{2\tau_d}{\tau_0}\right)\frac{\B
T^p_{dis}}{\tau_d}
\!-\!I_2\left(\frac{2\tau_d}{\tau_0}\right)
\left(\frac{\B T^p_{dis}}{\tau_d}:\B M \right)\frac{\B
T^p_{dis}}{\tau_d} \!-\!
\frac{\tau_0}{\tau_d}\ I_1\left(\frac{2\tau_d}{\tau_0}\right)
\B M\right],
\end{equation}
where $I_1(\cdot)$ and $I_2(\cdot)$ are the first- and second- order modified
Bessel functions of the first kind and $\tau_d\!=\!|\B T^p_{dis}|$ and $\tau_0\!\equiv\!2k_B \theta_k /v_0\gamma_s$. Expanding Eq. (\ref{simple_flow}) to first-order in
$\tau_d/\tau_0\!\ll\! 1$, we obtain
\begin{equation}
\label{D_1st} \B D^p \simeq \frac{2 v_0\gamma_s \nu}{\tau_0}\,\Lambda\, \,
e^{-\Delta/k_B\theta_k}\left(\B T^p_{dis}
-\tau_0 \B M\right) \ .
\end{equation}

This simplified flow rule has been used in the body of the manuscript for illustrative purposes without insisting on satisfying the condition $\tau_d/\tau_0\ll 1$. Next, we need to make sure that the flow rule in Eq. (\ref{D_1st}) does not violate the second law constraint in (\ref{plast_dissip}), which as explained previously, amounts to constraining ${\B T}^p_{store}$.
We then assume the following constitutive structure $\B T^p_{store}\!=\!\hat{C}(\Lambda,\theta_c,\theta_k)\B M$, i.e. that $\B T^p_{store}$ is linear in the orientation tensor $\B M$, which is then substituted together with $\B D^p$ of Eq. (\ref{D_1st}) in the inequality in (\ref{plast_dissip}). The resulting inequality is satisfied
if we set
\begin{equation}
\label{Tp_store_result}
\hat{C}(\Lambda,\theta_k,\theta_c)=\frac{2k_B\theta_c}{v_0\, \gamma_s} (s_0-1) \quad \Rightarrow \quad \B T^p_{store}=\frac{2 k_B\theta_c}{v_0\, \gamma_s} (s_0-1)\, \B M\ .
\end{equation}

The expression for $\B T^p_{store}$ in Eq. (\ref{Tp_store_result}) is interesting and deserves some discussion.
The prefactor relating $\B T^p_{store}$ and $\B M$ is proportional to $k_B \theta_c$, which implies that $\B T^p_{store}$
in this analysis is of {\em entropic} nature. More specifically, it suggests that configurational fluctuations -- whose magnitude is determined by $k_B\theta_c$ --
generate an entropic stress proportional to the orientation tensor $\B M$. Recalling the meaning of $s_0$ (defined in Eq. (\ref{m_entropy})), a lower bound value for $s_0$ is $(3/2)\log{2}$, obtainable by analyzing the minimal case where flow defects have two possible internal orientations.  Hence, $s_0\!>\!1$, which gives positivity of the prefactor in Eq. (\ref{Tp_store_result}).

\section{Frame-Indifference}
\label{frame_indifference}

Here we show the constitutive dependences discussed in Section \ref{constit_sys} are objective or frame-indifferent. We suppose a rigid-body change in frame of the observer,
\begin{equation}
\chib^*(\B{X}, t) = \B Q(t) ( \chib(\B X,t) -\B o)+ \B y(t)
\end{equation}
where $\B Q(t)$ is an arbitrary time-varying rotation about some origin $\B o$ and $\B y(t)$ is an arbitrary translation vector.  Hence, $\chib^*$ is the motion as observed by the new observer; we will continue to use $^*$ to describe quantities as seen in the new frame.

Frame-indifference is upheld if satisfaction of the constitutive functions in the original frame implies satisfaction in the new frame.  To show this, we first need to transform the field variables into their corresponding forms in the $^*$ frame and then check that the same constitutive functions correctly relate the transformed variables.

We start with the fundamental transformations:
\begin{equation}
\B F^*=\B Q (t) \B F\ , \ \ \ \B T^*=\B Q (t) \B T \B Q (t)^T \ .
\end{equation}
Under the Kr\"{o}ner-Lee kinematic decomposition, the $\B F$ transformation above is satisfied by letting
\begin{equation}
\B F^{e*}=\B Q (t)\B F^e\ , \ \ \ \B F^{p*}=\B F^p \
\end{equation}
(cf. \citet{Gurtin10}). Consequently,
\begin{equation}
\B L^{p*}=\dot{\B F}^{p*}\B F^{p*-1}=\dot{\B F}^{p}\B F^{p-1}=\B L^p \ \ \to \ \ \B D^{p*}=\B D^{p}, \  d^{p*}=d^p \ ,
\end{equation}
and
\begin{equation}
\B C^{e*}=\B F^{e*T}\B F^{e*}=\B F^{eT}\B Q (t)^T \B Q (t)\B F^e=\B C^e \ \ \to \ \ \B E^{e*}=\B E^e \ .
\end{equation}
Combining the latter expression with the transformation for $\B T$, the elastic and plastic stresses obey $\B T^{e*}=\B T^e\ , \ \ \B T^{p*}=\B T^p .$ The rate of stored plastic power $ \B T^p_{store}:\B L^p$ must, as a material scalar, be frame-invariant (i.e. $\B T^p_{store}:\B L^p= T^{p*}_{store}:\B L^{p*}$).  Since $\B L^p$ has just been shown to be frame-invariant as well, we deduce $\B T^{p*}_{store}=\B T^p_{store}$, and thus $\B T^{p*}_{dis}=\B T^{p*}-\B T^{p*}_{store}=\B T^{p}-\B T^{p}_{store}=\B T^p_{dis}$ .

The anisotropy internal variable naturally obeys $\B M^*=\B M$, since $\B M$ is defined in the intermediate (i.e. structural) space, which is the same in both frames according to $\B F^{p*}=\B F^p$ \ .

All scalar constitutive quantities (e.g. $\eta_c$, $\eta_k$, $\epsilon_c$, $\epsilon_k$, $\theta_c$, $\theta_k$, $\Lambda$, $q_{ck}$) are inherently frame-invariant. Combining this fact with the transformation rules above, we have now shown that the dependent and independent variables related under the constitutive functions $\hat{\eta}_k$, $\hat{\eta}_c$, $\hat{\B T}^e$, $\hat{\dot{{\Lambda}}}$, $\hat{\dot{{\B M}}}$, $\hat{\B T}^p_{dis}$, $\hat{\B T}^p_{store}$ proposed in Sections \ref{sec:conventional} and  \ref{constit_sys}, are all frame-invariant. Likewise, the functional dependences as proposed necessarily satisfy frame-indifference.

Lastly, we check that the heat-flux constitutive functions are also objective. Let us temporarily bring back the sub- and superscript $I$ to denote intermediate space variables (recall they were introduced and then dropped in Section \ref{math}).  The heat flux vector in the current configuration $\B q_k$ transforms under the rotating new frame as $\B q_k^*=\B Q(t)\B q_k$.  Pulled back to the intermediate space, we have
\begin{equation}
\B q_k^{I*}=J^{e*}\B F^{e*-1}\B q_k^*=J^{e} \B F^{e-1}\B Q(t)^T \B Q(t)\B q_k  = J^{e} \B F^{e-1}\B q_k=\B q_k^{I} \ .
\end{equation}
Similarly the temperature gradient $\B g_k=\nabla\theta_k$ transforms as $\B g_k^*=\B Q(t) \nabla \theta$.  Likewise
\begin{equation}
\B g_k^{I*}=\B F^{e*T}\nabla \theta_k^*=\B F^{eT}\B Q(t)^T \B Q(t) \nabla \theta_k=\B F^{eT}\nabla \theta_k=\B g_k^I \ .
\end{equation}
We thus have that $\B q_k^I$ and $\B g_k^I$ are both frame-invariant.  The same argument gives that $\B q_c^I$ and $\B g_c^I$ are also frame-invariant.  Consequently, all dependent and independent variables of the functions $\hat{\B q}_k$ and $\hat{\B q}_c$ in Section \ref{constit_sys} are frame-invariant, implying that the functional dependences are themselves indifferent to frame.

\end{document}